\documentclass[12pt]{article}
\begin{document}
%%%%%%%%%%%%%%%%%%%%%%%%%%%%%%%%
%\begin{center}
%{\Large {\bf Membrane and Noncommutativity}}
%\vskip 2cm
%{\bf Author, Author and Author}
%\vskip 2cm
%Address
%
%Address
%\end{center}
%%%%%%%%%%%%%%%%%%%%%%%%%%%%%%%%%
\title{{\bf Membrane and Noncommutativity}}

\author{
{\normalsize {\bf Rabin Banerjee\thanks{On leave of absence from S.~N.~Bose National Centre for Basic Sciences, Kolkata, India; \newline rabin@post.kek.jp, rabin@bose.res.in},}}\\
{\normalsize Institute of Particle and Nuclear Studies,}\\[-0.12cm]
{\normalsize High Energy Accelerator Research Organisation (KEK),}\\[-0.12cm]
{\normalsize Tsukuba 305-0801, Japan}
\and 
{\normalsize {\bf Biswajit Chakraborty\thanks{biswajit@bose.res.in} and Kuldeep Kumar\thanks{kuldeep@bose.res.in}}}\\
{\normalsize S.~N.~Bose National Centre for Basic Sciences,}\\[-0.12cm]
{\normalsize Block JD, Sector III, Salt Lake, Kolkata-700\,098, India}}

\date{{\normalsize \today}}

\maketitle

%%%%%%%%%%%%%%%%%%%%%%%%%%%%%%%%%%
\begin{abstract}
We analyse the dynamics of an open membrane, both for the free case and when it is coupled to a background three-form, whose boundary is attached to $p$-branes.  The role of boundary conditions and constraints in  the Nambu-Goto and Polyakov formulations is studied. The low-energy approximation that effectively reduces the membrane to an open string is examined in detail. Noncommutative features of the boundary string coordinates, where the cylindrical membrane is attached to the D$p$-branes, are revealed by algebraic consistency arguments and not by treating boundary conditions as primary constraints, as is usually done. The exact form of the noncommutative algebra is obtained in the low-energy limit.

{\bf Keywords:} Constrained hamiltonian analysis, Membranes, Noncommutativity

{\bf PACS:} 11.10.Nx, 11.25.-w, 11.25.-q, 11.10.Ef

\end{abstract}

%%%%%%%%%%%%%%%%%%%%%%%%%%%%%%%%%%%%%%%%
\section{Introduction}

Over the last decade string theory has been gradually replaced by M-theory as
the most natural candidate for a fundamental description of nature. While a
complete definition of M-theory is yet to be given, it is believed that the
five perturbatively consistent string theories are different phases of this
theory. With the replacement of string theory by M-theory, the string itself
has lost its position as the main candidate for  the fundamental degree of
freedom. Instead, higher-dimensional extended objects like membranes are being
considered \cite{jhwt}.
Indeed it is known that
membrane and five-brane occur naturally in eleven-dimensional supergravity,
which is
argued to be the low-energy limit of M-theory. Also, string theory is
effectively described by the low-energy dynamics of a system of branes. For
instance, the
membrane of M-theory may be ``wrapped" around the compact direction of radius
$R$ to become the fundamental string of type-IIA string theory, in the limit of
vanishing radius.

An intriguing connection between string theory, noncommutative geometry,
noncommutative (as well as ordinary) Yang-Mills theory was revealed
in \cite{sw}. With the shift in focus from string theory
to M-theory, there has been a flurry of activity in analysing noncommutativity
in membranes, specifically when an open membrane that couples to a three-form,
ends on a D-brane \cite{bbss,ks,dmm,t}.
The motivation of the present paper is to further this investigation, but with
a new perspective and methodology, as explained below.

The study of noncommutative properties in membranes is more involved than the
analogous study in the string case since the equations to be solved are
nonlinear. Naturally, in contrast to the string situation, the results could
be obtained only under some approximations. It is useful to recapitulate how
noncommutativity is derived in either the string coupled to the two-form or
the membrane coupled to the three-form. There are non-trivial boundary
conditions which are incompatible with the basic Poisson brackets of the
theory. These boundary conditions are considered as primary constraints in
the algorithm of Dirac's constrained hamiltonian dynamics
\cite{aas1, aas2, ks, dmm, t, ch, rv}. The primary
constraints lead to secondary constraints. Noncommutativity is manifested
through the occurrence of non-trivial Dirac brackets. The
brackets are found to be gauge dependent, but there is no gauge where it can be made to vanish.

Recently, an alternative approach to deal with noncommutativity in strings was
advocated in a paper \cite{bcg} involving  two of us. Contrary to other approaches,  the boundary conditions are not interpreted as primary constraints. The
noncommutative algebra emerges from a set of consistency requirements.  It is
rather similar in spirit to the original analysis of \cite{hrt} where a modified
algebra,  involving the periodic delta function instead of the usual one, was
found for
the coordinates and their conjugate momenta, in the example of the free
Nambu-Goto (NG) string.

In this paper we adopt our previous strategy for strings to the membrane model.
We discuss both the NG and Polyakov forms of action, although
noncommutativity is explicitly considered only in the latter formulation. The
similarities or otherwise in the analysis of the two actions are illuminated.
Analogous to the set of orthonormal gauge fixing conditions given for the free
Nambu-Goto string \cite{bcg, hrt}, we derive a set of quasi-orthonormal gauge conditions
for the free NG membrane.
Just as the orthonormal gauge in the NG string corresponds to the
conformal gauge in the Polyakov string, we find out the analogue of the
quasi-orthonormal gauge in the Polyakov membrane. It corresponds to a choice
of the metric that leads to equations of motion that can be explicitly solved
in the light-front
coordinates \cite{wt}. The structure and implications of the boundary conditions in the
two formulations have been elaborated. In the NG case, the conditions involve
the velocities that cannot be inverted so that a phase-space formulation is
problematic. Only by fixing a gauge is it possible to get hold of a phase-space description. In the Polyakov type, on the other hand, the boundary
condition is expressible
in phase-space variables without the need of any gauge choice.
This is because the metric itself is regarded as an independent field. In this
sense, therefore, there is no qualitative difference between string and
membrane
boundary conditions, since even in the NG string, a gauge fixing is required
for writing the boundary conditions in terms of phase-space variables.  We
thus differ from  \cite{dmm} where it is claimed that it is imperative in the
membrane case, as opposed to the string case, to gauge fix in order to express
the boundary conditions in phase-space coordinates as a first step in the
hamiltonian formalism.

The mandatory gauge fixing in the NG membrane, as  we shall show, converts
the reparametrization-invariant (first-class) system into a second-class one,
necessitating the use of Dirac brackets. This involves the inversion of
highly non-linear expressions, so that approximations become essential to make
any progress. Hence we avoid this formulation in favour of the Polyakov
version, where gauge fixing is not mandatory.

A detailed constrained hamiltonian analysis of the free bosonic Polyakov
membrane naturally
leads to three restrictions on the world-volume metric. These are found to be
identical to those obtained by counting the independent degrees of freedom.
Unlike the case of the classical string where there are three components of
the metric and three continuous symmetries (two diffeomorphism symmetries and
one scale symmetry), leading to a complete specification of the metric by
gauge fixing, for the membrane there are six independent metric components
and only three diffeomorphism symmetries. Thus only three restrictions on the
metric can be imposed.
Interestingly, the restrictions usually put in by hand \cite{wt} to perform
calculations in the light-front coordinates are obtained directly in our
hamiltonian formalism.  This gauge fixing is only partial in the sense that
the non-trivial gauge generating first-class constraints remain unaffected.
Effectively, therefore, it is a gauge-independent hamiltonian formalism. We
show that the boundary string coordinates corresponding to the membrane-D$p$-brane system (i.e. when the boundary of the open membrane is attached to
$p$-branes) satisfy the usual Poisson algebra without any noncommutativity.
By imposing further gauge conditions, it is possible to simulate a situation
where the cylindrical membrane is wrapped around a circle of vanishing radius
so that the open membrane passes over to an open string. The boundary
conditions of the membrane reduce to the well-known Neumann
boundary conditions of the string in the conformal gauge, just as the membrane
metric reduces to the conformal metric of the Polyakov string.

Next, the interacting membrane in the presence of a constant three-form tensor
potential is discussed. Proceeding in a gauge-independent manner, it is shown
that, contrary to the free theory, the boundary string coordinates must be
noncommutative. This is shown from certain algebraic conditions. However, in
contrast to the string case where it was possible to solve  these equations
\cite{bcg}, here an explicit solution is prevented from the non-linear
structure. Nevertheless, by passing to the low-energy limit (wrapping the
membrane on a circle of vanishingly small radius),
the explicit form of the noncommutativity in an open string, whose end points
are attached to a D-brane,  are reproduced.

The paper is organised as follows: In section~{\bf \ref{sec:ng}}, the free NG
membrane is
discussed and the form of the quasi-orthonormal gauge conditions, which act
as the analogue of the orthonormal gauge conditions in the NG string
\cite{hrt}, is derived. The role of the boundary conditions in maintaining stability of the membrane is discussed. The free Polyakov membrane is considered in
section~{\bf \ref{sec:poly}}, where its
detailed constrained hamiltonian account is given. The complete form of the
energy-momentum tensor is derived. All components of this tensor are written
as a linear combination
of the constraints. This is a generalization of the string case since even
though Weyl symmetry is absent in the membrane, the energy-momentum tensor
has a (weakly) vanishing trace; namely, it vanishes only on the constraint
shell. The brackets for the free theory with a cylindrical topology for the membrane, computed in
section~{\bf \ref{sec:brac}}, yield
the expected Poisson algebra without any noncommutativity. The low-energy limit
where the membrane is approximated by the string, is discussed
in section~{\bf \ref{sec:low}}.
Section~{\bf \ref{sec:int}} gives an analysis of the
interacting theory. General algebraic requirements enforce a noncommutativity
of the boundary coordinates of the membrane, which are attached to the
$p$-branes. No gauge fixing or approximation is needed to reveal this
noncommutativity. The explicit structure of the algebra is once again
computed in the low-energy approximation, when the result agrees with the
conformal-gauge expression for the noncommutativity among the coordinates of
the end points of the string attached to D-branes. Concluding remarks are
given in section~{\bf \ref{sec:conc}}. An appendix, summarizing the basic results of our earlier paper \cite{bcg} on strings, has been included for easy comparison with the membrane analysis.

%%%%%%%%%%%%%%%%%%%%%%%%%%%%%%%%%%%%%%%%
\section{The Free Nambu-Goto Membrane}
\label{sec:ng}

A dynamical membrane moving in $D-1$ spatial dimensions sweeps out a 
three-dimensional world-volume in $D$-dimensional space-time. We use a metric
with signature $(-, +, +, \cdots, +)$ in the target space whose indices are 
$\mu, \nu = 0, 1, 2, \ldots, (D-1)$. We can locally choose a set of three 
coordinates
$\sigma^{i}, i=0,1,2$, on the world-volume to parameterize it. We shall
sometime use the notation $\tau = \sigma^{0}$ and  
the indices $a, b, \ldots$ to describe ``spatial'' coordinates $\sigma^{a}, 
a = 1, 2$, on the membrane world-volume. In such a coordinate system, the
motion of the membrane through space-time is described by a set of $D$
functions $X^{\mu}(\sigma^{0}, \sigma^{1},\sigma^{2})$ which are the membrane
coordinates in the target space.

Although we are going to study the noncommutativity through the Polyakov 
action, we find it convenient to briefly discuss the NG action 
also. The NG analysis will be just an
extension of the string case, considered in \cite{hrt}. The NG action
for a membrane moving in flat space-time is given by the integrated proper 
volume swept out by the membrane:
\begin{equation}
S_{NG} = -T\int_{\Sigma} d^{3}\sigma \sqrt{-h} \equiv \int_{\Sigma} d^{3}\sigma {\mathcal L}_{NG}
\left(X^{\mu},\partial_{i}X^{\mu}\right),
\label{101}
\end{equation}
where $T$ is a constant which can be interpreted as the membrane tension and
$h = \det h_{ij}$ with
\begin{equation}
h_{ij} = \partial_{i}X^{\mu}\partial_{j}X_{\mu}
\label{102}
\end{equation}
being the induced metric on (2+1)-dimensional world-volume, which is nothing
but the pullback of the flat space-time metric on this three-dimensional 
sub-manifold. This induced metric, however, does not have the status of an
independent field in the world-volume; it is rather determined through the
embedding fields $X^{\mu}$.
The Lagrangian density is ${\mathcal L}_{NG} = -T\sqrt{-h}$.
The Euler-Lagrange equation is given by
\begin{equation}
\partial_{i}\left(\sqrt{-h}h^{ij}\partial_{j}X^{\mu}\right) = 0,
\label{103}
\end{equation}
while the boundary conditions are given by
\begin{equation}
\left.{\mathcal P}^{a}_{\mu}\right|_{\partial\Sigma}=\left.-T\sqrt{-h}
\partial^{a}X_{\mu}\right|_{\partial\Sigma} = 0,
\label{104}
\end{equation}
where
\begin{equation}
{\mathcal P}^{i}_{\mu} = \frac{\partial {\mathcal L}_{NG}}{\partial 
(\partial_{i}X^{\mu})} = -T\sqrt{-h}\partial^{i}X_{\mu}
\label{105}
\end{equation}
and $\partial \Sigma$ represents the boundary.
The components ${\mathcal P}^{0}_{\mu} \equiv \Pi_{\mu}$ are the canonical 
momenta conjugate to $X^{\mu}$. 
Using this, the Euler-Lagrange equation~(\ref{103}) can be rewritten as
\begin{equation}
\partial_{0}\Pi^{\mu}+\partial_{a}{\mathcal P}^{a\mu} = 0.
\label{106}
\end{equation}
It can be seen easily that the theory admits the
following primary constranits:
\begin{eqnarray}
\psi \!\!&\equiv&\!\! \Pi^{2} + T^{2} \overline{h} \approx 0,
\label{107} \\
\phi_{a} \!\!&\equiv&\!\! \Pi_{\mu}\partial_{a}X^{\mu} \approx 0, \quad a = 1,2,
\label{108}
\end{eqnarray}
where  $\Pi^{2} \equiv \Pi^{\mu}\Pi_{\mu}$ and $ \overline{h}=\det h_{ab}
= h_{11}h_{22}-(h_{12})^{2}$. These constraints are first-class since the
brackets between them vanish weakly:
\begin{eqnarray}
\left\{\psi\left(\tau,\vec{\sigma}\right),\psi\left(\tau,\vec{\sigma}'\right)\right\}
\!\!&=&\!\!4T^{2}\left[
\left\{h_{22}\left(\tau,\vec{\sigma}\right)\partial_{1}\delta\left(\vec{\sigma}\!-\!\vec{\sigma}'\right)-h_{12}\left(\tau,\vec{\sigma}\right)\partial_{2}\delta\left(\vec{\sigma}\!-\!\vec{\sigma}'\right)\right\}\phi_{1}\left(\tau,\vec{\sigma}\right)\right.
\nonumber \\
&\quad&\quad+\left\{h_{11}\left(\tau,\vec{\sigma}\right)\partial_{2}\delta\left(\vec{\sigma}\!-\!\vec{\sigma}'\right)-h_{12}\left(\tau,\vec{\sigma}\right)\partial_{1}\delta\left(\vec{\sigma}\!-\!\vec{\sigma}'\right)\right\}\phi_{2}\left(\tau,\vec{\sigma}\right)
\nonumber \\
&\quad&\quad-\left\{h_{22}\left(\tau,\vec{\sigma}'\right)\partial'_{1}\delta\left(\vec{\sigma}\!-\!\vec{\sigma}'\right)-h_{12}\left(\tau,\vec{\sigma}'\right)\partial'_{2}\delta\left(\vec{\sigma}\!-\!\vec{\sigma}'\right)\right\}\phi_{1}\left(\tau,\vec{\sigma}'\right)
\nonumber \\
&\quad&\quad\left.-\left\{h_{11}\left(\tau,\vec{\sigma}'\right)\partial'_{2}\delta\left(\vec{\sigma}\!-\!\vec{\sigma}'\right)-h_{12}\left(\tau,\vec{\sigma}'\right)\partial'_{1}\delta\left(\vec{\sigma}\!-\!\vec{\sigma}'\right)\right\}\phi_{2}\left(\tau,\vec{\sigma}'\right)\right]
\nonumber \\
\!\!&\approx&\!\!0,
\label{1109} \\
\left\{\phi_{a}\left(\tau,\vec{\sigma}\right),\phi_{b}\left(\tau,\vec{\sigma}'\right)\right\}
\!\!&=&\!\!\phi_{b}\left(\tau,\vec{\sigma}\right)\partial_{a}\delta\left(\vec{\sigma}\!-\!\vec{\sigma}'\right)-\phi_{a}\left(\tau,\vec{\sigma}'\right)\partial'_{b}\delta\left(\vec{\sigma}\!-\!\vec{\sigma}'\right)\approx0,
\nonumber \\
\left\{\psi\left(\tau,\vec{\sigma}\right),\phi_{a}\left(\tau,\vec{\sigma}'\right)\right\}
\!\!&=&\!\!2\psi\left(\tau,\vec{\sigma}\right)\partial_{a}\delta\left(\vec{\sigma}\!-\!\vec{\sigma}'\right)+\partial_{a}\psi\,\delta\left(\vec{\sigma}\!-\!\vec{\sigma}'\right)\approx0,
\nonumber
\end{eqnarray}
where $\partial'_{a}=\frac{\partial}{\partial \sigma'^{a}}$ and $\vec{\sigma}=(\sigma^{1},\sigma^{2})$.

The canonical world-volume energy-momentum tensor \emph{density}{\footnote{
Note that ${\mathcal L}_{NG}$ transforms as a scalar \emph{density} under
diffeomorphism.}}
can be obtained through
Noether theorem:
\begin{equation}
{[\theta_{C}]^{i}}_{j} = \frac{\partial{\mathcal L}_{NG}}{\partial(\partial_{i}
X^{\mu})}\partial_{j}X^{\mu}-{\delta^{i}}_{j}{\mathcal L}_{NG}.
\label{1110}
\end{equation}
In particular, ${[\theta_{C}]^{0}}_{0}=0$, ${[\theta_{C}]^{0}}_{a}
= \phi_{a}\approx0$, ${[\theta_{C}]^{a}}_{0}=0$ and ${[\theta_{C}]^{a}}_{b}=0$.
We notice that the canonical Hamiltonian density,
${\mathcal H}_{C}={[\theta_{C}]^{0}}_{0}$, obtained by Legendre
transformation, vanishes strongly.
Since the canonical energy-momentum tensor density is first-class, we may add
to it a 
linear combination of first-class constraints with tensor-valued coefficients 
to write down the total energy-momentum tensor density as
\begin{equation}
{\theta^{i}}_{j} = {U^{i}}_{j}\psi+{V^{ai}}_{j}\phi_{a} \approx 0.
\label{1111}
\end{equation}
The generators of $\tau$ and $\sigma^{a}$ translations are
\begin{equation}
H_{T}=\int\!\!d^{2}\sigma{\theta^{0}}_{0}, \quad 
H_{a}=\int\!\!d^{2}\sigma{\theta^{0}}_{a}.
\label{1112}
\end{equation}
As one can easily see, there are no secondary constraints. 
The Hamilton's equation $\dot{X}^{\mu} = \{X^{\mu}, H_{T}\}$ gives
\[
\partial_{0}X^{\mu} = 2{U^{0}}_{0}\Pi^{\mu}+{V^{a0}}_{0}\partial_{a}X^{\mu},
\]
which reproduces the definition of momenta $\Pi^{\mu}$ for the following
choice of ${U^{0}}_{0}$ and ${V^{a0}}_{0}$:
\begin{equation}
{U^{0}}_{0} = \frac{\sqrt{-h}}{2T\overline{h}},
\quad {V^{a0}}_{0} = -\frac{hh^{0a}}{\overline{h}}= \overline{h}^{ab}h_{0b},
\label{1113}
\end{equation}
where $\overline{h}^{ab}\, (\neq h^{ab}$, which is obtained by chopping off
first
row and first column from $h^{ij}$, the inverse of $h_{ij})$ is the inverse
of $h_{ab}$ in the two-dimensional subspace. The other equation,
$\dot{\Pi}^{\mu} = \{\Pi^{\mu}, H_{T}\}$, reproduces the Euler-Lagrange
equation~(\ref{103}) whereas $\partial_{a}X^{\mu} = \{X^{\mu}, H_{a}\}$ gives
\[
\partial_{a}X^{\mu} = 2{U^{0}}_{a}\Pi^{\mu}+{V^{b0}}_{a}\partial_{b}X^{\mu},
\]
which is satisfied for
\begin{equation}
{U^{0}}_{a}=0, \quad {V^{b0}}_{a}= {\delta^{b}}_{a}.
\label{1114}
\end{equation}

Coming to the conserved Poincar\'e generators in the target space, the translational generator is  
given by
\[
P_{\mu} = \int d^{2}\sigma \Pi_{\mu}, 
\]
and the angular momentum generator is given by
\[
M^{\mu\nu} = \int d^{2}\sigma\left(X^{\mu}\Pi^{\nu}-X^{\nu}\Pi^{\mu}\right).
\]
As can be easily checked, these generators generate appropriate Poincar\'e
transformations.
The above analysis can be generalized in a straightforward manner to an
arbitrary $p$-brane.

There is an interesting implication of the boundary conditions~(\ref{104}).
For a cylindrical membrane with $\sigma^{1} \in [0, \pi], \sigma^{2}
\in [0, 2\pi)$, $\sigma^{2}$ representing the compact direction,
the boundary condition is written as
\[
\left.\mathcal{P}^{1}_{\mu}\right|_{\sigma^{1}=0,\pi}=\left.-T\sqrt{-h}
\partial^{1}X_{\mu}\right|_{\sigma^{1}=0,\pi} = 0.
\]
Squaring the above equation, we get
\begin{equation}
\left.hh^{11}\right|_{\sigma^{1}=0,\pi}
= \left[h_{00}h_{22}-(h_{02})^{2}\right]_{\sigma^{1}=0,\pi} = 0,
\label{1123}
\end{equation}
which implies
\begin{equation}
h_{00}|_{\sigma^{1}=0,\pi} = \left.\frac{(h_{02})^{2}}{h_{22}}\right|_{\sigma^{1}=0,\pi}.
\label{112376}
\end{equation}
However, $h_{22}$ is strictly positive and cannot vanish at the boundary in
order to prevent it from collapsing to a point as the length of the boundary
is given by $\int^{2\pi}_{0}\sqrt{h_{22}}d\sigma^{2}$.
This indicates that
\[
\dot{X}^{2}|_{\sigma^{1}=0,\pi}=h_{00}|_{\sigma^{1}=0,\pi} \geq 0
\]
so that the points on the boundary move along either a space-like or light-like
trajectory. If we now demand that the speed of these boundary points should
not exceed the speed of light then we must have $h_{02}|_{\sigma^{1}=0,\pi}=0$
in Eq.~(\ref{112376}) so that
\[
\dot{X}^{2}|_{\sigma^{1}=0,\pi}=h_{00}|_{\sigma^{1}=0,\pi} = 0.
\]
Therefore the boundary points move with the speed of light which is a direct
generalization of the string case where a similar result holds.
For a square membrane with $\sigma^{1}, \sigma^{2} \in [0, \pi]$, the boundary
conditions~(\ref{104}) are written as
\begin{eqnarray*}
\left.\mathcal{P}^{1}_{\mu}\right|_{\sigma^{1}=0,\pi}\!\!&=&\!\!\left.-T\sqrt{-h}
\partial^{1}X_{\mu}\right|_{\sigma^{1}=0,\pi} = 0, \\
\left.\mathcal{P}^{2}_{\mu}\right|_{\sigma^{2}=0,\pi}\!\!&=&\!\!\left.-T\sqrt{-h}
\partial^{2}X_{\mu}\right|_{\sigma^{2}=0,\pi} = 0.
\end{eqnarray*}
Therefore, in addition to Eq.~(\ref{1123}), we also have
\[
\left.hh^{22}\right|_{\sigma^{2}=0,\pi}
= \left[h_{00}h_{11}-(h_{01})^{2}\right]_{\sigma^{2}=0,\pi} = 0.
\]
Proceeding just as in the case of cylindrical membrane, we find that we must
have $h_{02}|_{\sigma^{1}=0,\pi}=0$ and $h_{01}|_{\sigma^{2}=0,\pi}=0$ so that
\[
\dot{X}^{2}|_{\sigma^{1}=0,\pi} = 0 = \dot{X}^{2}|_{\sigma^{2}=0,\pi},
\]
which shows that the boundary points move with the speed of light. Also, since $h_{0a}\approx 0$
at the boundary, for both the cylindrical or square topology, it implies that the vector $\partial_0 X^\mu$ is not only null, but also orthogonal to all directions tangent to the membrane world-volume. Hence the boundary points move with the speed of light, perpendicularly to the membrane. This peculiar motion is exactly reminiscent of the string case. The tension in the free membrane would cause it to collapse. This is prevented by the angular momentum generated by the boundary motion, just as the collapse of the free string is thwarted by a similar motion of the string end points \cite{m}.

\subsection*{Quasi-orthonormal gauge fixing conditions}

As we shall see now, the membrane case, or any $p$-brane with $p > 1$
for that matter, involves some subtle issues. 
The first step is to provide  a set of complete 
gauge fixing conditions.  Taking a cue from the previous analysis  we would like to generalize the condition $h_{0a}\approx 0$, so that it holds everywhere, instead of just at the boundary. This is also quite similar in spirit to what is done for implementing the orthonormal gauge in the string case. Indeed, following the string analysis of  \cite{hrt}, we first impose the 
following gauge fixing conditions:
\begin{eqnarray}
\lambda_{\mu}\left(X^{\mu}\left(\tau, \vec{\sigma}\right)-\frac{P^{\mu}\tau}
{TA}\right) \approx 0,
\label{109} \\
\lambda_{\mu}\left(\Pi^{\mu}\left(\tau, \vec{\sigma}\right)-\frac{P^{\mu}}
{A}\right) \approx 0,
\label{110}
\end{eqnarray}
where $\lambda_{\mu}$ is an arbitrary constant vector and $A$ is taken to be the
``parametric area'' of the membrane. For example, if the membrane is of square
topology with $\sigma^{1}, \sigma^{2} \in [0, \pi]$, it will be $\pi^{2}$ and
for cylindrical topology with $\sigma^{1} \in [0, \pi], \sigma^{2}
\in [0, 2\pi)$ (membrane periodic along $\sigma^{2}$-direction), it will be 
$2\pi^{2}$. Clearly, this ``parametric area'' is not
an invariant quantity under two-dimensional diffeomorphism. One can think of
the square or cylindrical membrane to be flat at one instant to admit a
Cartesian-like coordinate system on the membrane surface which will provide a 
coordinate chart for it during its future time evolution.

Differentiating Eq.~(\ref{109}) with respect to $\tau$ and using
Eq.~(\ref{110}), we get
\begin{equation}
\lambda \cdot \dot{X} \approx \frac{\lambda \cdot P}{TA} 
\approx \frac{\lambda \cdot \Pi}{T}.
\label{111}
\end{equation}
Differentiating Eq.~(\ref{109}) with respect to $\sigma^{a}$, and
Eq.~(\ref{110}) with respect to $\tau$ we get
\begin{eqnarray}
\lambda \cdot \partial_{a}X \approx 0,
\label{112} \\
\partial_{0} \left(\lambda \cdot \Pi \right) \approx 0.
\label{113}
\end{eqnarray}
Using Eq.~(\ref{113}), it follows from the form (\ref{106}) of Euler-Lagrange
equation that
\begin{equation}
\partial_{a}\left(\lambda \cdot {\mathcal P}^{a}\right) \approx 0.
\label{114}
\end{equation}
Upon contraction with $\lambda^{\mu}$, the boundary conditions (\ref{104}) give
\begin{equation}
\left.\lambda \cdot {\mathcal P}^{a}\right|_{\partial\Sigma} = 0.
\label{115}
\end{equation}
Now we impose an additional gauge fixing condition\footnote{One can
generalize this gauge fixing condition~(\ref{116}) for higher-dimensional
hyper-membranes. Any $n$-dimensional divergenceless vector field $A^{a}$,
subjected to the boundary condition $A^{a}|_{\partial\Sigma}~=~0$ (just like
$\lambda \cdot {\mathcal P}^{a}$ in (\ref{114}) and (\ref{115})) can be
expressed as
$A^{a}=\varepsilon^{abc_{1}\ldots c_{n-2}}\partial_{b}B_{c_{1}\ldots c_{n-2}}$,
where $B_{c_{1}\ldots c_{n-2}}$ are the components of an $(n-2)$-form. Like
the Kalb-Ramond gauge fields, these $B$'s have a hierarchy of
``gauge symmetries'' given by $B~\rightarrow~B'~=~B~+~dB_{(n-3)}$,
$B_{(n-3)}~\rightarrow~B'_{(n-3)}~=~B_{(n-3)}~+~dB_{(n-4)}, \ldots$,
so on and so forth,
where $B_{(p)}$ is a $p$-form. 
One can therefore easily see that the demand $A^{a}~=~0$
entails $(n-1)$ additional constraints as there are $(n-1)$ independent
components of $B_{(n-2)}$. With two gauge fixing conditions of type
(\ref{109}) and (\ref{110}), this gives rise to $(n+1)$ number
 of independent constraints, which exactly matches with the number of
first-class constraints of the type~(\ref{107}) and (\ref{108})
of the theory. For the special case of $n~=~2$, $A^{a}~=~\varepsilon^{ab}\partial_{b}B$, where $B$ is now a pseudo-scalar. Clearly the demand $A^{a}~=~0$
is equivalent to the gauge fixing condition~(\ref{116}). For the case $n~=~3$,
$A^{a}~=~\varepsilon^{abc}\partial_{b}B_{c}$ so that 3-vector is expressed as
a curl of another 3-vector, in a standard manner, having only two transverse
degrees of freedom; the
longitudinal one having been eliminated through the above mentioned gauge
transformation.}
\begin{equation}
\varepsilon^{ab} \partial_{a}\left(\lambda \cdot {\mathcal P}_{b}\right) 
\approx 0.
\label{116}
\end{equation}
Thus, we have
from Eqs.~(\ref{114}) and (\ref{116}) both the divergence and curl vanishing for
the vector field $(\lambda \cdot {\mathcal P}^{a})$ in the 2-dimensional
membrane, which is also subjected to the boundary conditions (\ref{115}).
We thus have
\begin{equation}
\lambda \cdot {\mathcal P}^{a} = 0 \quad \forall\,\, \sigma^{a}.
\label{117}
\end{equation}
In view of Eq.~(\ref{112}), we have
\[
\lambda \cdot \Pi \approx -T\sqrt{-h}h^{00}(\lambda \cdot \partial_{0}X),
\]
which, using Eq.~(\ref{111}) gives
\begin{equation}
h^{00}\sqrt{-h} \approx -1.
\label{118}
\end{equation}
Using Eqs.~(\ref{111}) and (\ref{112}), Eq.~(\ref{117}) gives $h^{0a} \approx 0$
which implies
\begin{equation}
h_{0a} \approx 0, \quad h^{00} \approx \frac{1}{h_{00}}.
\label{119}
\end{equation}
From Eqs.~(\ref{118}) and (\ref{119}) it follows that
\begin{equation}
h_{00}+\overline{h} \approx 0.
\label{120}
\end{equation}
Observe that the term quasi-orthonormality in this case means that the 
time-like vector $\partial_{0}$ is orthogonal to the 
space-like vectors
$\partial_{a}$, which follows from Eq.~(\ref{119}).
However, the two space-like directions $\partial_{1}$
and $\partial_{2}$ need not be orthogonal to each 
other. Also note that by replacing $\tau \rightarrow \alpha \tau$, $\alpha$ a
constant number, in Eq.~(\ref{109}), the normalization condition (\ref{120})
will change to $h_{00}+\alpha^{2}\overline{h} \approx 0$.

Using the quasi-orthonormal conditions~(\ref{119}) and (\ref{120}), the
Lagrangian density becomes
\[
\mathcal{L}_{NG} \approx -T\overline{h} \approx Th_{00}
\approx \frac{T}{2}\left(h_{00}-\overline{h}\right).
\]
The effective action thus becomes
\begin{equation}
S^{\mathrm{eff}} = \frac{T}{2}\int_{\Sigma} d^{3}\sigma\left[h_{00}-h_{11}h_{22}+(h_{12})^{2}\right],
\label{12006081}
\end{equation}
which gives the equation of motion:
\begin{equation}
\partial_{0}\partial_{0}X_{\mu} + \partial_{1}\left(h_{12}\partial_{2}X_{\mu}-h_{22}\partial_{1}X_{\mu}\right) + \partial_{2}\left(h_{12}\partial_{1}X_{\mu}-h_{11}\partial_{2}X_{\mu}\right) = 0.
\label{12006082}
\end{equation}

Note that the quasi-orthonormal conditions~(\ref{119}) and (\ref{120}) do
not correspond to any gauge conditions themselves as they contain time 
derivatives. Actually they follow as a consequence of the 
conditions~(\ref{109}), (\ref{110}) and (\ref{116}) which are to be regarded 
as gauge fixing conditions. These gauge conditions, when imposed, render
the first-class constraints~(\ref{107}) and (\ref{108}) of the theory into
second-class as can be seen from their non-vanishing Poisson-bracket structure.
Therefore, NG formalism requires the evaluation of Dirac brackets where these
constraints are implemented strongly. As we shall see subsequently, in the
Polyakov formulation the constraints~(\ref{107}) and (\ref{108}) are not 
rendered into second-class and we can avoid the detailed calculation of 
Dirac brackets.

It is possible to draw a parallel between the quasi-orthonormal gauge discussed here and the usual orthonormal gauge in NG string, which is the analogue of the conformal gauge in the Polyakov string. In the latter case the equations of motion linearize reducing to the D'Alembert equations. This is possible because the gauge choice induces a net of coordinates that form a locally orthonormal system \cite{r}. For the membrane, the invariances are insufficient to make such a choice and the best that we could do was to provide a quasi-orthonormal system. It is however amusing to note that if we forced an orthonormal choice, so that $h_{0a}\approx 0$ is supplemented with $h_{12}\approx 0$ and $h_{11}=h_{22}\approx 1$, then the equation of motion (\ref{12006082}) indeed simplifies to the D'Alembert equation. This provides an alternative way of looking at the quasi-orthonormality.

If we do not impose quasi-orthonormality,
it is highly non-trivial, if not totally impossible, to express 
the boundary conditions~(\ref{104})
in terms of phase-space variables because the canonical momentum
 $\Pi_{\mu}~=~\mathcal{P}^{0}_{\mu}$
(\ref{105}), which can
be re-expressed as
\[
\Pi_{\mu} = \frac{T\overline{h}}{\sqrt{-h}}\left(\eta_{\mu\nu}-\partial_{a}
X_{\mu}\overline{h}^{ab}\partial_{b}X_{\nu}\right)\partial_{0}X^{\nu}
\]
involves a projection operator given by the expression within the parentheses
in the above equation. The velocity terms appear both in the right of the 
projection operator and in $\sqrt{-h}$ appearing in the denominator. This
makes the inversion of the above equation to write the velocities in terms 
of momenta highly non-trivial. Nevertheless, all this simplifies drastically
in the quasi-orthonormal gauge to enable us to simplify the above expression to
\begin{equation}
\Pi_{\mu} = T\partial_{0}X_{\mu}
\label{12006084}
\end{equation}
so that the boundary condition~(\ref{104}) is now expressible in terms of phase-space variables
as
\[
\left.\left(h_{22}\partial_{1}X_{\mu}-h_{12}\partial_{2}X_{\mu}\right)\Pi^{2}\right|_{\sigma^{1}=0,\pi}=0.
\]

Finally we notice that the parameters ${U^{0}}_{0}$ and ${V^{a0}}_{0}$
given by Eq.~(\ref{1113}) simplify in this gauge to
\begin{equation}
{U^{0}}_{0} = \frac{1}{2T}, \quad {V^{a0}}_{0} = 0
\label{12006083}
\end{equation}
while ${U^{0}}_{a}$ and ${V^{b0}}_{a}$ given by Eq.~(\ref{1114}) remain
unchanged.
Now the generators of $\tau$ and $\sigma^{a}$ translations (\ref{1112})
become
\begin{equation}
H_{T}=\frac{1}{2T}\int\!\!d^{2}\sigma \psi, \quad 
H_{a}=\int\!\!d^{2}\sigma \phi_{a}.
\label{121}
\end{equation}
It is straightforward to reproduce the action (\ref{12006081}) by performing an inverse Legendre transformation.
Computing the Poisson bracket of $X_{\mu}(\tau, \vec{\sigma})$ with the above
$H_{T}$, the Hamilton's equation $\partial_{0}X_{\mu} = \{X_{\mu}, H_{T}\}$
gives Eq.~(\ref{12006084}), the definition of momenta in this gauge. Then,
\[
S^{\mathrm{eff}}= \int_\Sigma d^3\sigma \,\Pi_\mu\partial_0 X^\mu - \int d\tau\, H_T
\]
just yields (\ref{12006081}). 
The other equation,
$\partial_{0}\Pi_{\mu} = \{\Pi_{\mu}, H_{T}\}$, reproduces
Eq.~(\ref{12006082}),
which is the Euler-Lagrange equation following from the effective
action~(\ref{12006081}).

Notice that the values of ${U^{0}}_{0}$ and ${V^{a0}}_{0}$ are gauge dependent.
The particular values given by Eq.~(\ref{12006083}) correspond to our
quasi-orthonormal gauge.
Had we chosen a different gauge, we would have obtained different values for
these parameters. On the contrary, the parameters
${U^{0}}_{a}$ and ${V^{b0}}_{a}$ are gauge independent. This is consistent with the symmetries of the problem. There are three reparametrization invariances, so that three parameters among these $U$'s and $V$'s must be gauge dependent, manifesting these symmetries. Since the reparametrization invariances govern the time evolution of the system,  the gauge-dependent parameters are given by ${U^{0}}_{0}$ and ${V^{a0}}_{0}$, while the others are gauge independent.

%%%%%%%%%%%%%%%%%%%%%%%%%%%%%%%%%%%%%%%%
\section{The Free Polyakov Membrane}
\label{sec:poly}

The Polyakov action for the bosonic membrane is \cite{wt}
\begin{equation}
S_{P} = -\frac{T}{2}\int_{\Sigma} d^{3}\sigma \sqrt{-g}\left(g^{ij}\partial_{i}X^{\mu}
\partial_{j}X_{\mu} - 1\right),
\label{301}
\end{equation}
where an auxiliary metric $g_{ij}$ on the membrane world-volume has been
introduced and will be given the status of an independent field variable in
the enlarged configuration space.
The final term $(-1)$ inside the parentheses
does not appear in the analogous string theory action.
A consistent set of equations can be obtained only by taking the ``cosmological'' constant to be $(-1)$. Indeed, the equations of motion following from  the action (\ref{301}) but with arbitrary cosmological constant $\Lambda$ are
\begin{eqnarray}
\partial_{i}\left(\sqrt{-g}g^{ij}\partial_{j}X^{\mu}\right) = 0,
\label{302} \\
h_{ij} = \frac{1}{2}g_{ij}\left(g^{kl}h_{kl}+\Lambda\right)
\label{303}
\end{eqnarray}
while the boundary conditions are 
\begin{equation}
\left.\partial^{a}X^{\mu} \right|_{\partial \Sigma} = 0.
\label{304}
\end{equation}
Eq.~(\ref{303}) can now be satisfied if and only if we identify $g_{ij}$ with
$h_{ij}$:
\begin{equation}
g_{ij} = h_{ij} \equiv \partial_{i}X^{\mu}\partial_{j}X_{\mu}
\label{305}
\end{equation}
for the case $\Lambda = -1$ so that the action (\ref{301}) reduces to the NG action (\ref{101}).
The canonical momenta corresponding to the fields $X^{\mu}$ and $g_{ij}$ are
\begin{eqnarray}
\Pi_{\mu} \!\!&=&\!\! \frac{\partial {\mathcal L}}{\partial \dot{X}^{\mu}}
= -T\sqrt{-g}\partial^{0}X_{\mu},
\label{306} \\
\pi^{ij} \!\!&=&\!\! \frac{\partial {\mathcal L}}{\partial \dot{g}_{ij}} = 0.
\label{307}
\end{eqnarray}
Clearly, $\pi^{ij} \approx 0$ represent primary constraints of the theory.
The canonical Hamiltonian density is 
\begin{eqnarray}
{\mathcal H}_{C}\!\!&=&\!\!\Pi_{\mu}\partial_{0}X^{\mu} - {\mathcal L}
\nonumber \\
\!\!&=&\!\! \frac{\sqrt{-g}}{2T\overline{g}}\Pi^{2}
- \frac{gg^{0a}}{\overline{g}}\Pi_{\mu}\partial_{a}X^{\mu}
+ \frac{T\sqrt{-g}}{2\overline{g}}\left(g_{22}h_{11}
+g_{11}h_{22}-2g_{12}h_{12}-\overline{g}\right).
\label{308}
\end{eqnarray}
Therefore, the total Hamiltonian is written as
\begin{equation}
H_{T} = \int d^{2}\sigma \left({\mathcal H_{C}}+\lambda_{ij}\pi^{ij}\right),
\label{309}
\end{equation}
where $\lambda_{ij}$ are arbitrary Lagrange multipliers.
Conserving the constraint $\pi^{00}\approx 0$ with time:
\[
\dot{\pi}^{00} = \left\{\pi^{00}, H_{T}\right\} \approx 0,
\]
we get
\begin{equation}
\Omega_{1} \equiv \Pi^{2} + T^{2}\left( g_{22}h_{11} + g_{11}h_{22} - 2g_{12}h_{12}
- \overline{g}\right) \approx 0.
\label{3010}
\end{equation}
Similarly, conserving other primary constraints with time, we get
\begin{eqnarray}
\Omega_{2}\!\!&\equiv&\!\!\frac{\sqrt{-g}}{4T{\overline{g}}^{2}}
\left(2g_{22}-\overline{g}g^{11}\right)\left\{\Pi^{2}+ T^{2}\left( g_{22}h_{11} + g_{11}h_{22} - 2g_{12}h_{12}\right)\right\}
\nonumber \\
&&\!\!- \frac{gg_{22}}{\overline{g}^{2}}g^{0a}\Pi_{\mu}\partial_{a}X^{\mu}
- \frac{g_{02}}{\overline{g}}\Pi_{\mu}\partial_{2}X^{\mu}
- \frac{T\sqrt{-g}}{4\overline{g}}\left(2h_{22}-\overline{g}g^{11}\right)
\approx 0,
\label{3011} \\
\Omega_{3}\!\!&\equiv&\!\!\frac{\sqrt{-g}}{4T{\overline{g}}^{2}}
\left(2g_{11}-\overline{g}g^{22}\right)\left\{\Pi^{2}+ T^{2}\left( g_{22}h_{11} + g_{11}h_{22} - 2g_{12}h_{12}\right)\right\}
\nonumber \\
&&\!\!- \frac{gg_{11}}{\overline{g}^{2}}g^{0a}\Pi_{\mu}\partial_{a}X^{\mu}
- \frac{g_{01}}{\overline{g}}\Pi_{\mu}\partial_{1}X^{\mu}
- \frac{T\sqrt{-g}}{4\overline{g}}\left(2h_{11}-\overline{g}g^{22}\right)
\approx 0,
\label{3012} \\
\Omega_{4}\!\!&\equiv&\!\!-\frac{\sqrt{-g}}{2T{\overline{g}}^{2}}
\left(2g_{12}+\overline{g}g^{12}\right)\left\{\Pi^{2}+ T^{2}\left( g_{22}h_{11} + g_{11}h_{22} - 2g_{12}h_{12}\right)\right\}
\nonumber \\
&&\!\!+ \frac{2gg_{12}}{\overline{g}^{2}}g^{0a}\Pi_{\mu}\partial_{a}X^{\mu}
+ \frac{g_{02}}{\overline{g}}\Pi_{\mu}\partial_{1}X^{\mu}
+ \frac{g_{01}}{\overline{g}}\Pi_{\mu}\partial_{2}X^{\mu}
+ \frac{T\sqrt{-g}}{2\overline{g}}\left(2h_{12}+\overline{g}g^{12}\right)
\approx 0,
\label{3013} \\
\Omega_{5}\!\!&\equiv&\!\!-\frac{\sqrt{-g}g^{01}}{2T}\left\{\Pi^{2}+ T^{2}\left( g_{22}h_{11} + g_{11}h_{22} - 2g_{12}h_{12}-\overline{g}\right)\right\}
\nonumber \\
&&\!\!- g_{22}\Pi_{\mu}\partial_{1}X^{\mu}
+ g_{12}\Pi_{\mu}\partial_{2}X^{\mu}
\approx 0,
\label{3014} \\
\Omega_{6}\!\!&\equiv&\!\!-\frac{\sqrt{-g}g^{02}}{2T}\left\{\Pi^{2}+ T^{2}\left( g_{22}h_{11} + g_{11}h_{22} - 2g_{12}h_{12}-\overline{g}\right)\right\}
\nonumber \\
&&\!\!- g_{11}\Pi_{\mu}\partial_{2}X^{\mu}
+ g_{12}\Pi_{\mu}\partial_{1}X^{\mu}
\approx 0.
\label{3015}
\end{eqnarray}
The above constraints appear to have a complicated form. Also, their connection with the constraints obtained in the NG formalism, is not particularly transparent. To bring the constraints into a more tractable form and to illuminate this connection, it is desirable to express them by the following combinations:
\begin{eqnarray}
\Omega_{1}\!\!&=&\!\!\psi - T^{2}\overline{\chi} \approx 0,
\label{30101} \\
\Omega_{2}\!\!&=&\!\!\frac{\sqrt{-g}}{4T{\overline{g}}^{2}}
\left(2g_{22}-\overline{g}g^{11}\right)\Omega_{1}
- \frac{gg_{22}}{\overline{g}^{2}}g^{0a}\phi_{a}
- \frac{g_{02}}{\overline{g}}\phi_{2}
+ \frac{T\sqrt{-g}}{2\overline{g}}\chi_{22}
\approx 0,
\label{30112} \\
\Omega_{3}\!\!&=&\!\!\frac{\sqrt{-g}}{4T{\overline{g}}^{2}}
\left(2g_{11}-\overline{g}g^{22}\right)\Omega_{1}
- \frac{gg_{11}}{\overline{g}^{2}}g^{0a}\phi_{a}
- \frac{g_{01}}{\overline{g}}\phi_{1}
+ \frac{T\sqrt{-g}}{2\overline{g}}\chi_{11}
\approx 0,
\label{30123} \\
\Omega_{4}\!\!&=&\!\!-\frac{\sqrt{-g}}{2T{\overline{g}}^{2}}
\left(2g_{12}+\overline{g}g^{12}\right)\Omega_{1}
+ \frac{2gg_{12}}{\overline{g}^{2}}g^{0a}\phi_{a}
+ \frac{g_{02}}{\overline{g}}\phi_{1}
+ \frac{g_{01}}{\overline{g}}\phi_{2}
- \frac{T\sqrt{-g}}{\overline{g}}\chi_{12}
\approx 0,
\label{30134} \\
\Omega_{5}\!\!&=&\!\!-\frac{\sqrt{-g}g^{01}}{2T}\Omega_{1}
- g_{22}\phi_{1} + g_{12}\phi_{2}
\approx 0,
\label{30145} \\
\Omega_{6}\!\!&=&\!\!-\frac{\sqrt{-g}g^{02}}{2T}\Omega_{1}
- g_{11}\phi_{2} + g_{12}\phi_{1}
\approx 0,
\label{30156}
\end{eqnarray}
where
\begin{eqnarray}
\psi \!\!&\equiv&\!\! \Pi^{2} + T^{2}\overline{h} \approx 0,
\label{310} \\
\phi_{a} \!\!&\equiv&\!\! \Pi_{\mu}\partial_{a}X^{\mu} \approx 0,
\label{311} \\
\chi_{ab} \!\!&\equiv&\!\! g_{ab}-h_{ab} \approx 0
\label{312} 
\end{eqnarray}
and $\overline{\chi}=\chi_{11}\chi_{22}-(\chi_{12})^{2}$.
As all the constraints $\Omega$'s appearing in Eqs.~(\ref{30101}--\ref{30156})
are combinations of $\psi$, $\phi_{a}$ and $\chi_{ab}$ in
Eqs.~(\ref{310}--\ref{312}), we can treat these
$\psi$, $\phi_{a}$ and $\chi_{ab}$ as an alternative set of secondary
constraints.
These constraints along with the primary constraints
$\pi^{ij} \approx 0$ (\ref{307}) constitute the
complete set of constraints of the theory. This is because the canonical
Hamiltonian density (\ref{308}) can be expressed as a combination of
constraints in the following manner:
\begin{equation}
{\cal H}_{C} = \frac{\sqrt{-g}}{2T\overline{g}}\psi - \frac{gg^{0a}}
{\overline{g}}\phi_{a} - \frac{T\sqrt{-g}}{2\overline{g}}\overline{\chi}
\approx 0
\label{313}
\end{equation}
and the non-vanishing Poisson brackets between the constraints of the theory
are
\begin{eqnarray}
\left\{\psi(\tau, \vec{\sigma}), \chi_{ab}(\tau, \vec{\sigma}')\right\}
\!\!&\approx&\!\! 2\left(\partial_{a}\Pi_{\mu}\partial_{b}X^{\mu}+\partial_{b}
\Pi_{\mu}\partial_{a}X^{\mu}\right)\delta\left(\vec{\sigma}- \vec{\sigma}'
\right),
\nonumber \\
\left\{\phi_{a}(\tau, \vec{\sigma}), \chi_{bc}(\tau, \vec{\sigma}')\right\}
\!\!&=&\!\! h_{ab}(\tau, \vec{\sigma}')\partial'_{c}\delta\left(\vec{\sigma}
-\vec{\sigma}'\right) + h_{ac}(\tau, \vec{\sigma}')\partial'_{b}\delta
\left(\vec{\sigma}- \vec{\sigma}'\right) 
\nonumber \\
&\quad& + \left(\partial_{b}X^{\mu}\partial_{c}\partial_{a}X_{\mu}
+\partial_{c}X^{\mu}\partial_{b}\partial_{a}X_{\mu}\right)\delta
\left(\vec{\sigma}-\vec{\sigma}'\right),
\label{314} \\
\left\{\pi^{ab}(\tau, \vec{\sigma}), \chi_{cd}(\tau, \vec{\sigma}')\right\}
\!\!&=&\!\! -\frac{1}{2}\left(\delta^{a}_{c}\delta^{b}_{d}+\delta^{a}_{d}\delta^{b}_{c}\right)\delta\left(\vec{\sigma}- \vec{\sigma}'\right),
\nonumber 
\end{eqnarray}
while the
weakly vanishing brackets are the same as given by (\ref{1109}).
As far as the rest of the brackets are concerned, it is trivial to see that
they vanish strongly. Thus, as it appears, none of the constraints except
$\pi^{0i}$ in the set is first-class. But we have not yet extracted the
maximal number of first-class constraints from the set (\ref{307},
\ref{310}--\ref{312}) by constructing appropriate linear combinations of the
constraints. However, it is highly non-trivial to find such a linear
combination in the present case as one can see from the complicated structure
of the Poisson brackets given above in (\ref{314}).
Nevertheless, one can bypass such an elaborate procedure to extract the 
first-class constraints from the given set by noting that the complete set of constraints can be split into two sectors. In one sector we retain $\psi, \phi_a $ and $\pi^{0i}$, which are first-class among themselves, while the other sector contains the canonically conjugate pairs $\chi_{ab}$ and 
$\pi^{ab}$.  This allows an iterative computation of the Dirac brackets \cite{gt}; namely, it is possible to eliminate this set completely by calculating the Dirac brackets within this sector. The brackets of the other constraints are now computed with respect to these Dirac brackets. Obviously $\psi, \phi_a$ will have vanishing brackets with $\pi^{ab}, \chi_{cd}$. Moreover, the original first-class algebra among $\psi$ and $\phi_a$ will be retained. This follows from the fact that  the Dirac constraint matrix involving $\pi^{ab}$ and $\chi_{cd}$ has entries only in the off-diagonal pieces, while $\psi$ and $\phi_a$ have non-vanishing contributions coming just from the bracket with one of them; i.e. $\chi_{cd}$ (see (\ref{314})). 
The Dirac brackets of $\psi$ and $\phi_{a}$ are thus identical to their 
Poisson brackets, satisfying the same algebra as in the NG case.

We are therefore left with the first-class constraints $\psi \approx 0$,
$\phi_{a} \approx 0$ and $\pi^{0i} \approx 0$. At this stage, we note that the
constraints $\pi^{0i} \approx 0$ are analogous to $\pi^{0} \approx 0$ in free
Maxwell theory, where $\pi^{0}$ is canonically conjugate to $A_{0}$.
Consequently, the time evolution of $g_{0i}$ is arbitrary as follows from the
Hamiltonian (\ref{309}). Therefore, we can set
\begin{equation}
g_{0a} = 0, \quad g_{00} = -\overline{h},
\label{316}
\end{equation}
as new gauge fixing conditions.\footnote{We cannot set $g_{00} = 0$ as it will
make the metric singular. We therefore set $g_{00} = -\overline{h}$ to make it
match with the corresponding condition (\ref{120}) in NG case.} With that
($g_{0a},\pi^{0a}$) and
($g_{00},\pi^{00}$) are discarded from the phase-space. This is again
analogous to the arbitrary time evolution of $A_{0}$ in Maxwell theory, where
we can set $A_{0}=0$ as a gauge fixing condition and discard the pair
($A_{0},\pi^{0}$) from the phase-space altogether.

These gauge fixing conditions (\ref{316}) are the counterpart of the
quasi-orthonormal conditions (\ref{119}) and (\ref{120}) in the
NG case. However, unlike the NG case, these second-class constraints
(\ref{316}) do not
render the residual first-class constraints of the theory, viz.
$\psi \approx 0$ and $\phi_{a} \approx 0$ into second-class constraints.
Therefore, they represent partial gauge fixing conditions.
This stems from the fact that $g_{0i}$ were still regarded as independent field
variables in the configuration space whereas $g_{ab}$ have already been
strongly 
identified with $h_{ab}$ (\ref{312}). We therefore note that the
calculation of the Dirac brackets is not
necessary in Polyakov formulation. This motivates us to study the
noncommutativity vis-\`{a}-vis the modified brackets
$\{X^{\mu},X^{\nu}\}$ in the simpler Polyakov version. For that we shall first
consider the free theory in the next section.

Before we conclude this section, let us make some pertinent observations about
the structure of the symmetric form of energy-momentum tensor, which is
obtained by functionally differentiating the action with respect to the metric. The various components of this tensor are given by:
\begin{eqnarray}
T_{00} \!\!&=&\!\! \frac{g_{00}}{2T\overline{g}}\psi+\frac{2g\sqrt{-g}}{{\overline{g}^{2}}}g^{0a}\phi_{a}+\left(\frac{1}{g^{00}}-g_{00}\right)\frac{\Pi^{2}}{T\overline{g}} 
\nonumber \\
\!\!&\quad&\!\!-\frac{Tg^{2}}{\overline{g}^{2}}\left[(g^{01})^{2}h_{11}+(g^{02})^{2}h_{22}+2g^{01}g^{02}h_{12}\right]-\frac{Tg_{00}}{2\overline{g}}\overline{\chi},
\label{317} \\
T_{01}\!\!&=&\!\!-\frac{g_{01}}{2T\overline{g}}\psi-\frac{\sqrt{-g}}{\overline{g}}\phi_{1}+\frac{T}{\overline{g}}\left[\left(g_{02}h_{11}+g_{01}h_{12}\right)\chi_{12}-g_{02}h_{12}\chi_{11}-g_{01}h_{11}\chi_{22}-\frac{g_{01}}{2}\overline{\chi}\right], \quad
\label{318} \\
T_{02}\!\!&=&\!\!-\frac{g_{02}}{2T\overline{g}}\psi-\frac{\sqrt{-g}}{\overline{g}}\phi_{2}+\frac{T}{\overline{g}}\left[\left(g_{01}h_{22}+g_{02}h_{12}\right)\chi_{12}-g_{01}h_{12}\chi_{22}-g_{02}h_{22}\chi_{11}-\frac{g_{02}}{2}\overline{\chi}\right],
\label{319} \\
T_{ab}\!\!&=&\!\!-\frac{g_{ab}}{2T\overline{g}}\psi+T\chi_{ab}+\frac{Tg_{ab}}
{2\overline{g}}\overline{\chi}-\frac{Tg_{ab}}{\overline{g}}\left(g_{22}\chi_{11}+g_{11}\chi_{22}-2g_{12}\chi_{12}\right).
\label{320}
\end{eqnarray}
Note that unlike the case of string \cite{bcg}, the component $T_{00}$ cannot be
written in terms of constraints of the theory. However, the other
components can be expressed in terms of these constraints, of which
$\chi_{ab}$ are second-class and have already been put strongly to  zero by
using Dirac brackets, so that the form of $T_{0a}$ and $T_{ab}$ simplifies to
\begin{eqnarray*}
T_{0a}\!\!&=&\!\!-\frac{g_{0a}}{2T\overline{g}}\psi-\frac{\sqrt{-g}}{\overline{g}}\phi_{a}, \\
T_{ab}\!\!&=&\!\!-\frac{g_{ab}}{2T\overline{g}}\psi.
\end{eqnarray*}
However, for $T_{00}$ we have to make use of the gauge conditions (\ref{316}), which hold strongly as was discussed earlier, to enable us to write
\[
T_{00}=-\frac{1}{2T}\psi.
\]
Let us now compare it with NG case. First we notice that ${\theta^{i}}_{j}$ appearing in Eq.~(\ref{1111}) is not a tensor itself but it is a tensor density. The corresponding tensor is $\frac{1}{\sqrt{-g}}{\theta^{i}}_{j}$. In quasi-orthonormal gauge, we have
\[
\sqrt{-g}{T^{0}}_{0}={\theta^{0}}_{0}=\frac{1}{2T}\psi,
\]
which reproduces the canonical Hamiltonian density (\ref{313}) in this
gauge. Also, in this gauge, we have
\[
\sqrt{-g}{T^{0}}_{a}=\phi_{a},
\]
which matches with ${\theta^{0}}_{a}$ in quasi-orthonormal gauge.
This also provides a direct generalization of the string case \cite{bcg}.
Although,
unlike the string case, the Weyl symmetry is absent in the membrane case, we
still have a vanishing trace, albeit weakly, of the energy-momentum tensor:
\[
{T^{i}}_{i} = - \frac{1}{2T\overline{h}}\psi \approx 0.
\]

%%%%%%%%%%%%%%%%%%%%%%%%%%%%%%%%%%%%%%%%
\section{The Brackets for a Free Theory}
\label{sec:brac}

Here we consider a cylindrical topology for the membrane which is taken to be
periodic along $\sigma^{2}$-direction, i.e. $\sigma^{2}\in [0, 2\pi)$ and
$\sigma^{1} \in [0, \pi]$. Following the example of string case \cite{bcg},
we write down the first version of the brackets as:
\begin{equation}
\{X^{\mu}(\tau ,\vec{\sigma}),\Pi_{\nu}(\tau ,\vec{\sigma}')\}
= \delta^{\mu}_{\nu}\Delta_{+}(\sigma^{1}, \sigma'^{1})\delta_{P}(\sigma^{2}\!-\!\sigma'^{2}),
\label{401} 
\end{equation}
and the other brackets vanishing.\footnote{Note that the
$\{X^{\mu},\Pi_{\nu}\}$ brackets are not affected as we implemented the second-class constraints and the gauge fixing conditions strongly in the preceding section. They are the only surviving phase-space variables as $g_{ij}$ have lost their independent status.} Here
\begin{equation}
\delta_{P}(\sigma-\sigma')={1\over {2\pi}}\sum_{n\in {\cal Z}}
e^{in(\sigma-\sigma')}
\label{403}
\end{equation}
is the periodic delta function of period $2\pi$ which satisfies
\begin{equation}
\int_{-\pi}^{+\pi}\!d\sigma'\delta_{P}(\sigma-\sigma')f(\sigma')=f(\sigma)
\label{404}
\end{equation}
for any periodic function $f(\sigma)=f(\sigma+2\pi )$ defined in the interval
$[-\pi, +\pi ]$; and if, in addition, $f(\sigma)$ is taken to be an even
function in the interval $[-\pi, +\pi ]$, then the above integral (\ref{404})
reduces to
\begin{equation}
\int_0^{\pi}d\sigma'\Delta_{+}(\sigma,\sigma')f(\sigma')=f(\sigma),
\label{405}
\end{equation}
where
\begin{eqnarray}
\Delta_{+}(\sigma,\sigma') \!\!&=&\!\! \delta_{P}(\sigma-\sigma')+\delta_{P}(\sigma+\sigma')
\nonumber \\
&=&\!\! {1\over \pi }+{1\over \pi }\sum_{n \neq 0}\cos(n\sigma)\cos(n\sigma').
\label{406}
\end{eqnarray}
This structure of the brackets is, however, consistent only with Neumann
boundary conditions along $\sigma^{1}$-direction. On the other hand, we have
a mixed boundary condition (\ref{304}) which can be expressed in terms of
phase-space variables as
\begin{equation}
\left[g_{22}T\partial_{1}X^{\mu} + \sqrt{-g}g^{01}\Pi^{\mu}
- g_{12}T\partial_{2}X^{\mu}\right]_{\sigma^{1}=0,\pi} = 0.
\label{407}
\end{equation}
We notice that in NG formulation it was necessary to fix gauge in order to
express the
boundary condition in terms of phase-space variables. However, this is not the
case with Polyakov formulation since $g_{ij}$ are taken to be independent
fields.
Using the strongly valid equations~(\ref{312}) and the gauge fixing
conditions~(\ref{316}), this simplifies further to
\begin{equation}
\left[\partial_{2}X^{\nu}\partial_{2}X_{\nu}\partial_{1}X^{\mu}
- \partial_{1}X^{\nu}\partial_{2}X_{\nu}\partial_{2}X^{\mu}
\right]_{\sigma^{1}=0,\pi} = 0.
\label{408}
\end{equation}
Although we are using the gauge~(\ref{316}), the non-trivial gauge generating
first-class constraints~(\ref{310}) and (\ref{311}) will be retained in the
gauge-independent analysis both here and in the interacting case. As we see,
the above boundary condition is non-trivial in nature and involves both the
$\partial_{1}$ and $\partial_{2}$ derivatives. But, since the coordinates and
momenta are not related
at the boundary, we do not require to postulate a non-vanishing
$\{X^{\mu}, X^{\nu}\}$ bracket as in the case of free string in conformal
gauge \cite{bcg}. Therefore, the free membrane theory, like its string
counterpart, does not exhibit noncommutativity in the boundary coordinates.

%%%%%%%%%%%%%%%%%%%%%%%%%%%%
\section{The Low-Energy Limit}
\label{sec:low}

In this section, we would like to see how the results in the
free membrane theory go over to those of free string theory in the limit of 
small radius for the cylindrical membrane.

The cylindrical membrane is usually taken to propagate in an 11-dimensional
compactified target space ${\mathbf R}^{9-p} \times M^{p} \times S^{1} \times
I$, where $M^{p}$ is a $p$-dimensional flat Minkowski space-time and $I$ is an
interval with finite length. There exist at the boundaries of $I$ two
$p$-branes on which an open membrane can end. And the topology of the 
$p$-branes is given by $M^{p} \times S^{1}$. Also, the cylindrical membrane is
assumed to wrap around this $S^{1}$. The radius of this circle is supposed to
be very small so that in the low-energy limit the target space effectively
goes over to 10-dimensional  ${\mathbf R}^{9-p} \times M^{p} \times I$ and
the cylindrical membrane goes over to the open string.

At this stage, we choose further gauge fixing conditions:
\begin{equation}
X^{0} = \tau, \quad X^{2} = \sigma^{2}R,
\label{410}
\end{equation}
where we have introduced $R$ to indicate the radius of the cylindrical
membrane and $X^{2}$ represents the compact dimension $S^{1}$.\footnote{In \cite{ks}, another gauge fixing condition
$X^{1}=\sigma^{1}$, ($\pi$ being the length of the cylindrical membrane) has
been used. But we notice that imposition of this gauge fixing condition would
be inconsistent with the boundary condition~(\ref{408}) since, for
$\mu = 1$, it yields a topology changing condition
(cylinder $\rightarrow$ sphere), $R^{2}|_{\sigma^{1}=0,\pi}=0$, which is 
clearly unacceptable.
Therefore, the
choice~(\ref{410}) does not allow us to choose $X^{1}=\sigma^{1}$ as well,
which is not needed either for our purpose.}
Before choosing the gauge conditions~(\ref{410}), the $\tau$ and $\sigma^{a}$
translations were generated by the constraints $\frac{1}{2T}\psi$ and $\phi_{a}$
respectively, just as in the NG case (\ref{121}). Now we have
\begin{eqnarray*}
\left\{\psi\left(\tau, \vec{\sigma}\right), X^{0}\left(\tau, \vec{\sigma}'\right)\!-\!\tau\right\}
\!\!&=&\!\! -2\Pi^{0}\left(\tau, \vec{\sigma}\right)\Delta_{+}(\sigma^{1},\sigma'{^1})\delta_{P}(\sigma^{2},\sigma'{^2}),
\\
\left\{\phi_{2}\left(\tau, \vec{\sigma}\right), X^{2}\left(\tau, \vec{\sigma}'\right)\!-\!\sigma'^{2}R\right\}
\!\!&=&\!\! -\partial_{2}X^{2}\left(\tau, \vec{\sigma}\right)\Delta_{+}(\sigma^{1},\sigma'{^1})\delta_{P}(\sigma^{2},\sigma'{^2})
\\
\!\!&\approx&\!\! -R\Delta_{+}(\sigma^{1},\sigma'{^1})\delta_{P}(\sigma^{2},\sigma'{^2}),
\end{eqnarray*}
whereas
\begin{eqnarray*}
\left\{\phi_{1}\left(\tau, \vec{\sigma}\right), X^{0}\left(\tau, \vec{\sigma}'\right)\!-\!\tau\right\}
\!\!&=&\!\! -\partial_{1}X^{0}\left(\tau, \vec{\sigma}\right)\Delta_{+}(\sigma^{1},\sigma'{^1})\delta_{P}(\sigma^{2},\sigma'{^2})
\approx 0,
\\
\left\{\phi_{1}\left(\tau, \vec{\sigma}\right), X^{2}\left(\tau, \vec{\sigma}'\right)\!-\!\sigma'^{2}R\right\}
\!\!&=&\!\! -\partial_{1}X^{2}\left(\tau, \vec{\sigma}\right)\Delta_{+}(\sigma^{1},\sigma'{^1})\delta_{P}(\sigma^{2},\sigma'{^2})
\approx 0.
\end{eqnarray*}
Thus, the (partial) gauge fixing conditions~(\ref{410}) take care of the
world-volume diffeomorphism generated by $\psi$ and $\phi_{2}$ in the sense
that these constraints are rendered into second-class while
the diffeomorphism generated by $\phi_{1}$ is still there.

Coming back to the low-energy limit, we would like to show that the
$\sigma^{2}$ dependence of all the fields except $X^{2}$ itself drops out
effectively in the gauge (\ref{410}). To motivate it, let us consider the
case of a free massless scalar field defined on
a space with one compact dimension of ignorable size.
Let the space be $M^{p} \times S^{1}$, where $M^{p}$ is
a $p$-dimensional Minkowski space-time taken to be flat for simplicity
and $S^{1}$ is a
circle of radius $R$ which is very small.
We take $\theta \in [0,2\pi)$ to be the angle
coordinate coorresponding to this circle so that the metric is given by
$ds^{2}=\eta_{\mu\nu}dx^{\mu}dx^{\nu}=\eta_{\mu'\nu'}dx^{\mu'}dx^{\nu'}+R^{2}d\theta^{2}$ with $\mu, \nu$ ranging from 0 to $p$ and $\mu', \nu'$ from 0 to
$(p-1)$. The action is
\[
S = -\frac{1}{2}\int\!\!d^{p}x\,d\theta\,\partial_{\mu}\phi\partial^{\mu}\phi.
\]
Separating the index corresponding to the compact dimension, we rewrite it as
\[
S=-\frac{1}{2}\int\!\!d^{p}x\,d\theta \left(\partial_{\mu'}\phi\partial^{\mu'}\phi+\frac{1}{R^{2}}\partial_{\theta}\phi\partial_{\theta}\phi\right).
\]
Substituting the Fourier expansion
\[
\phi(x,\theta)
= {1\over {\sqrt{2 \pi}}}\sum_{n\in {\cal Z}}\phi_{(n)}(x)e^{in\theta},
\quad \phi_{(-n)}=\phi_{(n)}^{*}
\]
in the action and integrating out the compact dimension, we get
\[
S\rightarrow S'= -\frac{1}{2}\int\!\!d^{p}x\sum_{n\in {\cal Z}}\left(\partial_{\mu'}\phi_{(n)}\partial^{\mu'}\phi^{*}_{(n)}+\frac{n^{2}}{R^{2}}\phi_{(n)}\phi^{*}_{(n)}\right).
\]
Thus the Fourier coefficients represent a whole tower of
effective massive complex scalar fields of mass $\sim \frac{n}{R}$ in a
lower-dimensional non-compact space-time. These masses are usually of the
Planck order if $R$ is of the order of Planck length and are therefore ignored
in the low-energy regime. Equivalently, one ignores the $\theta$ dependece of
the field $\phi$. This can also be understood from physical considerations.
In the low-energy limit, the associated wavelengths are very large as compared
to $R$ so that variation of the field along the circle is ignorable.

Now the membrane goes over to string in the low-energy regime when
the circle $S^{1}$ effectively disappears in the limit $R \rightarrow 0$.
So the field theory living in the membrane world-volume is expected to
correspond to the field theory living on string world-sheet. To verify this,
let us substitute the Fourier expansion of the world-volume fields
$X^{\mu}(\tau, \sigma^{1}, \sigma^{2})$ around $\sigma^{2}$:
\begin{equation}
X^{\mu}(\tau, \sigma^{1}, \sigma^{2})
= {1\over {\sqrt{2 \pi}}}\sum_{n\in {\cal Z}}X^{\mu}_{(n)}(\tau, \sigma^{1})e^{in\sigma^{2}}, \quad X^{\mu}_{(-n)}={X^{\mu}_{(n)}}^{*}
\label{414}
\end{equation}
in the Poisson bracket~(\ref{401}) to find that the Fourier coefficients
$X^{\mu}_{(n)}(\tau, \sigma^{1})$ satisfy
\begin{equation}
\{X^{\mu}_{(n)}(\tau, \sigma^{1}),\Pi_{\nu}^{(m)}(\tau, \sigma'^{1})\}
= \delta^{\mu}_{\nu}\delta^{-m}_{n}\Delta_{+}(\sigma^{1},\sigma'^{1}).
\label{413}
\end{equation}
As in the case of free scalar field discussed above, the Fourier
coefficients $X^{\mu}_{(0)}(\tau, \sigma^{1})$ will
represent the effective (real) fields in the string world-sheet satisfying
\begin{equation}
\{X^{\mu}_{(0)}(\tau, \sigma^{1}),\Pi_{\nu}^{(0)}(\tau, \sigma'^{1})\}
= \delta^{\mu}_{\nu}\Delta_{+}(\sigma^{1},\sigma'^{1}),
\label{415}
\end{equation}
which reproduces the Poisson bracket for string (see appendix).
The sub/superscript (0) will be dropped now onwards for convenience. Using
$\partial_{2}X^{\mu}=R\delta^{\mu}_{2}$, the boundary condition~(\ref{408})
gives 
\begin{equation}
\left.\partial_{1}X^{\mu}\right|_{\sigma^{1}=0,\pi} = 0
\label{411}
\end{equation}
so that we recover the boundary condition for free string in conformal gauge
(see appendix).\footnote{Actually, we do not get (\ref{411}) directly, rather
it is accompanied by a pre-factor $R^{2}$. However, this equation is not
satisfied trivially if $R\rightarrow 0$, as this limit should not be taken 
literally in a mathematical sense. This just means that $R$ should be taken
to have a very small non-zero value and presumably should be of the order of
Planck length, as we have mentioned earlier.}

Now we would like to show how
the gauge fixed world-volume membrane metric~(\ref{316}) reduces to the
world-sheet string metric in conformal gauge. For that first note that
the components of the metric tensor in a matrix form can be written as
\[
\{g_{ij}\}=\pmatrix{g_{00} & g_{01} & g_{02}
\cr g_{01} & g_{11} & g_{12}
\cr g_{02} & g_{12} & g_{22}}
=\pmatrix{g_{00} & 0 & 0
\cr 0 & h_{11} & 0
\cr 0 & 0 & R^{2}},
\]
where we have made use of the first gauge fixing condition in (\ref{316}) and
by now the strongly valid equations~(\ref{312}). Clearly, this matrix becomes
singular in the limit $R \rightarrow 0$ taken in a proper mathematical sense.
It must therefore correspond to a two-dimensional surface embedded in 
three-dimensional world-volume. The metric corresponding to it can be easily
obtained by
chopping off the last row and last column in the above three-dimensional
metric to get $\pmatrix{g_{00} & 0 \cr 0 & h_{11}}$.
Now, we make use of the second gauge fixing condition in (\ref{316}) to replace
$g_{00}$ by $(-\overline{h})$. However, this $\overline{h}$ can be simplified
further using the gauge~(\ref{410}) to get $R^{2}h_{11}$ so that the above
$2 \times 2$ matrix becomes $h_{11}\pmatrix{-R^{2} & 0 \cr 0 & 1}$ and
the diagonal elements get identified up to
a scale factor. It can now be put in the standard form, ${\rm diag}\,(-1,1)$, upto an
overall Weyl factor, by replacing the second condition in (\ref{316}) by
$g_{00}=-\alpha^{2} \overline{h}$ and choosing $\alpha$ suitably.
We also notice that using $h_{0a}=0$, the NG action for the membrane becomes
\[
S_{NG} = -T\int d^{3}\sigma \sqrt{-h_{00}\overline{h}},
\]
which using the gauge conditions~(\ref{410}) and integrating out $\sigma^{2}$, reduces to the NG action for string in orthonormal gauge:
\[
S_{NG} \rightarrow S'_{NG} = -2\pi RT\int d^{2}\sigma \sqrt{-h_{00}h_{11}}.
\]
This also shows that the string tension is $\sim TR$ if the original membrane
tension is given by $T$.  Actually one takes the limit $R\rightarrow 0$ together with the membrane tension $T\rightarrow \infty $ in such a way that their product $(TR)$ is finite. Such a limit was earlier discussed, from other considerations, in \cite{l}.

%%%%%%%%%%%%%%%%%%%%%%%%%%%%%%%%%%%%%%%%
\section{The Interacting Membrane}
\label{sec:int}

The Polyakov action for a  membrane moving in the presence of a constant
antisymmetric background field $A_{\mu\nu\rho}$ is
\begin{equation}
S_{P} = -\frac{T}{2}\int_{\Sigma} d^{3}\sigma \left[\sqrt{-g}\left(g^{ij}
\partial_{i}X^{\mu}\partial_{j}X_{\mu}
-1\right)+\frac{e}{3}\varepsilon^{ijk}\partial_{i}X^{\mu}\partial_{j}X^{\nu}
\partial_{k}X^{\rho}A_{\mu\nu\rho}\right],
\label{501}
\end{equation}
where we have introduced a coupling constant $e$.\footnote{As it stands,
the interaction term involving the three-form field
$A_{\mu\nu\rho}$ in (\ref{501}) is not gauge invariant under the transformation
$A \rightarrow A + d\Lambda$, where $\Lambda$ is a two-form field. One can,
however, make it gauge invariant by adding a surface term
$2e\int_{\partial\Sigma}B$,
where $B$ is a two-form undergoing the compensating gauge transformation
$B \rightarrow B - \Lambda$. But, using Stoke's theorem, this gets combined
to a single integral over the world-volume as $\int_{\Sigma}(A+dB)$ so that
$(A+dB)$ is gauge invariant as a whole. In the action (\ref{501}), $A$ is taken
to correspond to this gauge invariant quantity by absorbing $dB$ in $A$.}
The equations of motion are
\begin{eqnarray}
\partial_{i}\left(\sqrt{-g}g^{ij}\partial_{j}X_{\mu}+\frac{e}{2}
\varepsilon^{ijk}\partial_{j}X^{\nu}\partial_{k}X^{\rho}A_{\mu\nu\rho}\right)
= 0,
\label{502} \\
g_{ij} = h_{ij} \equiv \partial_{i}X^{\mu}\partial_{j}X_{\mu}.
\qquad \qquad \qquad 
\label{503}
\end{eqnarray}
Note that the second equation does not change from the free case $(e=0)$
despite the presence of interaction term as this term is topological in nature
and does not involve the metric $g_{ij}$. The canonical momenta are
\begin{eqnarray}
\Pi_{\mu} \!\!&=&\!\! \frac{\partial {\mathcal L}}{\partial \dot{X}^{\mu}}
= -T\left(\sqrt{-g}\partial^{0}X_{\mu}+\frac{e}{2}\varepsilon^{ab}
\partial_{a}X^{\nu}\partial_{b}X^{\rho}A_{\mu\nu\rho}\right),
\label{504} \\
\pi^{ij} \!\!&=&\!\! \frac{\partial {\mathcal L}}{\partial \dot{g}_{ij}} = 0.
\label{505}
\end{eqnarray}
For convenience, we define
\begin{equation}
\widetilde{\Pi}_{\mu} \equiv \Pi_{\mu}+\frac{eT}{2}\varepsilon^{ab}\partial_{a}X^{\nu}\partial_{b}X^{\rho}A_{\mu\nu\rho}
= -T\sqrt{-g}\partial^{0}X_{\mu}.
\label{506}
\end{equation}
Proceeding just as in the free case, the structure of the Hamiltonian density
${\mathcal H}_{C}$ and the set of constraints is obtained just by replacing
${\Pi}_{\mu} \rightarrow \widetilde{\Pi}_{\mu}$, so that we are finally left
with the following first-class constraints:
\begin{eqnarray}
\psi \!\!&\equiv&\!\! \widetilde{\Pi}^{2} + T^{2}\overline{h} \approx 0,
\label{507} \\
\phi_{a} \!\!&\equiv&\!\! \widetilde{\Pi}_{\mu}\partial_{a}X^{\mu} \approx 0
\label{508}
\end{eqnarray}
and, as argued in the free case, we adopt the same gauge fixing conditions
(\ref{316}).

For a cylindrical membrane periodic along $\sigma^{2}$-direction
with $\sigma^{1} \in [0, \pi], \sigma^{2}
\in [0, 2\pi)$, the boundary condition is given by
\begin{equation}
\left[\sqrt{-g}\partial^{1}X_{\mu}+e\partial_{2}X^{\nu}\partial_{0}X^{\rho}
A_{\mu\nu\rho}\right]_{\sigma^{1}=0,\pi} = 0,
\label{509}
\end{equation}
which when expressed in terms of phase-space variables looks as
\begin{equation}
\left[g_{22}T\partial_{1}X_{\mu}-g_{12}T\partial_{2}X_{\mu}
+\sqrt{-g}g^{01}\Pi_{\mu}+e\left(\Pi^{\rho}+eT\partial_{1}X^{\lambda}
\partial_{2}X^{\kappa}{A^{\rho}}_{\lambda\kappa}\right)
\partial_{2}X^{\nu}A_{\mu\nu\rho}
\right]_{\sigma^{1}=0,\pi} = 0.
\label{510}
\end{equation}
As in the free case, here also we use the strongly valid equations~(\ref{312})
and
the gauge fixing conditions~(\ref{316}) so that the above boundary condition
simplifies to
\begin{equation}
\left[T\partial_{2}X^{\nu}\partial_{2}X_{\nu}\partial_{1}X_{\mu}
-T\partial_{1}X^{\nu}\partial_{2}X_{\nu}\partial_{2}X_{\mu}
+ e\left(\Pi^{\rho}+eT\partial_{1}X^{\lambda}
\partial_{2}X^{\kappa}{A^{\rho}}_{\lambda\kappa}\right)
\partial_{2}X^{\nu}A_{\mu\nu\rho}
\right]_{\sigma^{1}=0,\pi} = 0.
\label{511}
\end{equation}
Here we notice that the above boundary condition involves both phase-space
coordinates $(X^{\mu}, \Pi_{\nu})$. Using the brackets of the free theory to
compute the Poisson bracket of the left-hand side of above equation with
$X_{\sigma}(\tau, \vec{\sigma}')$, we find that it does not vanish. The
boundary condition is therefore not compatible with the brackets of the
free theory. Thus, we have to postulate a 
non-vanishing $\{X^{\mu}, X^{\nu}\}$ bracket.{\footnote {In the case of free
Polyakov string also, the incompatibility of the boundary condition with the
basic Poisson brackets forces us to postulate  a non-vanishing $\{X^{\mu},X^{\nu}\}$. However, in contrast to the interacting string, this bracket vanishes in a particular gauge --- the conformal gauge. (See appendix.)}} For that we make an ansatz:
\begin{eqnarray}
\left\{X_{\mu}(\tau ,\vec{\sigma}),X_{\nu}(\tau ,\vec{\sigma}')\right\}
\!\!&=&\!\! \mathcal{C}_{\mu \nu}(\vec{\sigma},\vec{\sigma}')
\nonumber \\
\!\!&=&\!\! C_{\mu \nu}(\sigma^{1},\sigma'^{1})\delta_{P}(\sigma^{2}
-\sigma'^{2})
\label{512}
\end{eqnarray}
with
\begin{equation}
C_{\mu \nu}(\sigma^{1},\sigma'^{1})
=-C_{\nu \mu}(\sigma'^{1},\sigma^{1}).
\label{513}
\end{equation}
and the $\{X^{\mu},\Pi_{\nu}\}$ bracket is taken to be the same as in the free
case --- Eq.~(\ref{401}). At this stage, we note that the boundary
condition~(\ref{511}), if bracketted with $X_{\sigma}(\tau, \vec{\sigma}')$,
yields at the boundary
\begin{eqnarray}
\left[T\partial_{2}X^{\nu}\partial_{2}X_{\nu}\delta_{\mu}^{\lambda}
- T\partial_{2}X^{\lambda}\partial_{2}X_{\mu}
+ e^{2}T\partial_{2}X^{\kappa}\partial_{2}X^{\nu}A_{\mu\nu\rho}{A^{\rho\lambda}}_{\kappa}\right]\partial_{1}{\mathcal{C}}_{\lambda\sigma}
(\vec{\sigma},\vec{\sigma}') \qquad \quad &&
\nonumber \\
+\left[2T\partial_{1}X_{\mu}\partial_{2}X^{\kappa}-T\partial_{1}X^{\nu}\partial_{2}X_{\nu}\delta_{\mu}^{\kappa}-T\partial_{2}X_{\mu}\partial_{1}X^{\kappa}+e\Pi_{\rho}{A_{\mu}}^{\kappa\rho}\right. \qquad \quad \qquad \quad \quad \qquad &&
\nonumber \\
+\left.e^{2}T\partial_{1}X^{\lambda}\partial_{2}X^{\nu}\left(A_{\mu\nu\rho}{{A^{\rho}}_{\lambda}}^{\kappa}+{{A_{\mu}}^{\kappa}}_{\rho}{A^{\rho}}_{\lambda\nu}\right) \right]\partial_{2}{\mathcal{C}}_{\kappa\sigma}
(\vec{\sigma},\vec{\sigma}') \qquad \quad &&
\nonumber \\
= e\partial_{2}X^{\nu}A_{\mu\nu\sigma}\Delta_{+}(\sigma^{1}, \sigma'^{1})\delta_{P}(\sigma^{2}\!-\!\sigma'^{2}), \!\!\!&&
\label{514}
\end{eqnarray}
which involves both $\partial_{1}{\mathcal{C}}$ and
$\partial_{2}{\mathcal{C}}$ and leads to a contradiction if we put
${\mathcal{C}}_{\mu\nu}(\vec{\sigma},\vec{\sigma}')=0$. This is another way
of seeing that there must be a noncommutativity in the membrane coordinates.
However, there is no contradiction with
${\mathcal{C}}_{\mu\nu}(\vec{\sigma},\vec{\sigma}')~=~0$ provided
$A_{\mu\nu\rho}~=~0$, thereby implying that there is no noncommutativity in the
free theory.

Because of the non-linearity in the above equation, it is problematic to find an exact solution. It should however be stressed that the above relation has been derived in a general (gauge-independent) manner. At this point there does not seem to be any compelling reason to choose a particular gauge to simplify this equation further to enable an exact solution. Nonlinearity would, in all probability, prevent this.  This is in contrast to the string case where the analysis naturally leads to a class of light-cone gauges where the corresponding equation was solvable \cite{bcg}.
However, by 
taking recourse to the
low-energy approximation, we show that the results for the string case are recovered. To this end, we substitute the
expansion~(\ref{414}) in 
(\ref{512}) to get
\begin{equation}
\{X^{\mu}_{(n)}(\tau, \sigma^{1}),X^{\nu}_{(m)}(\tau, \sigma'^{1})\}
= \delta_{n,-m}C^{\mu\nu}(\sigma^{1},\sigma'^{1}).
\label{5131}
\end{equation}
But again, as in the free case, we retain only the real fields
$X^{\mu}_{(0)}(\tau, \sigma^{1})\equiv X^{\mu}(\tau, \sigma^{1})$ when we
consider the low-energy regime.
Using the gauge fixing conditions~(\ref{410}), the boundary condition~(\ref{511}) reduces to
\begin{equation}
\left[(TR)\partial_{1}X_{\mu}-e\Pi^{\rho}A_{\mu\rho2}-e^{2}(TR)\partial_{1}X_{\lambda}{A^{\rho\lambda}}_{2}A_{\mu\rho2}\right]_{\sigma^{1}=0,\pi}=0.
\label{515}
\end{equation}
Here, $X^{\mu}$ and $\Pi_{\nu}$ can be taken to correspond to
$X^{\mu}_{(0)}(\tau, \sigma^{1})$ and $\Pi_{\nu}^{(0)}(\tau, \sigma^{1})$
respectively. Thus we recover the boundary condition of the string theory in
conformal gauge with the correspondence $TR\leftrightarrow T_{S}$ and
$A_{\mu\nu2}\leftrightarrow B_{\mu\nu}$, where $T_{S}$ is the effective
(string) tension and $B_{\mu\nu}$ is the 2-form
background field appearing in the string theory \cite{bcg}. Now taking the
Poisson bracket of the boundary condition~(\ref{515}) with
$X_{\sigma}(\tau, \sigma^{1})$, the low-energy effective real fields, one
gets for $\mu \neq 2$ the following differential condition satisfied by
$C_{\mu\sigma}$ at the boundary
\begin{equation}
\left.T_{S}\left(\delta_{\mu}^{\lambda}-e^{2}A_{\mu\rho2}{A^{\rho\lambda}}_{2}\right)\partial_{1}C_{\lambda\sigma}(\sigma^{1},\sigma'^{1})
\right|_{\sigma^{1}=0,\pi}
= \left.eA_{\sigma\mu2}\Delta_{+}(\sigma^{1},\sigma'^{1})
\right|_{\sigma^{1}=0,\pi},
\label{516}
\end{equation}
which just reproduces the corresponding equation in string theory (see
appendix, in particular equation (\ref{449})). We therefore obtain the noncommutativity:
\begin{equation}
C_{\mu\nu}(\sigma^{1},\sigma'^{1})
={1\over 2}(NM^{-1})_{(\nu \mu )}[\Theta(\sigma^{1},\sigma'^{1})-\Theta(\sigma'^{1},\sigma^{1})]
+{1\over 2}(NM^{-1})_{[\nu \mu]}[\Theta(\sigma^{1},\sigma'^{1})+\Theta(\sigma'^{1},\sigma^{1})-1],
\label{517}
\end{equation}
where
\[
N_{\nu\sigma}=eA_{\nu\sigma2}, \quad {M^{\lambda}}_{\mu}=T_{S}\left(\delta^{\lambda}_{\mu}-e^{2}A_{\mu\rho2}{A^{\rho\lambda}}_{2}\right)
\]
with $(NM^{-1})_{(\nu \mu)}$ the symmetric and $(NM^{-1})_{[\nu \mu]}$ the
antisymmetric part of $(NM^{-1})_{\nu \mu}$ while 
\begin{equation}
\Theta(\sigma^{1},\sigma'^{1})={\sigma^{1} \over \pi}+{1\over \pi }
\sum_{n\neq 0}{1\over n}\sin(n\sigma^{1})\cos(n\sigma'^{1})
\label{518}
\end{equation}
being the generalized step function which satisfies
\begin{equation}
\partial_{1}\Theta(\sigma^{1},\sigma'^{1})=\Delta_{+}(\sigma^{1},\sigma'^{1}).
\label{519}
\end{equation}
It has the property
\begin{eqnarray*}
\Theta(\sigma^{1},\sigma'^{1})\!\!&=&\!\!1\quad{\rm for}\quad\sigma^{1}>\sigma'^{1}, \\
\Theta(\sigma^{1},\sigma'^{1})\!\!&=&\!\!0\quad{\rm for}\quad\sigma^{1}<\sigma'^{1}.
\end{eqnarray*}

%%%%%%%%%%%%%%%%%%%%%%%%%%%%%%%%%%%%%%%%
\section{Concluding Remarks}
\label{sec:conc}

We have analysed an open membrane, with square and cylindrical topology, ending on $p$-branes. Both the free case as well as the theory where the membrane is coupled to a background three-form potential were considered. 

For the free theory, the world-volume action was taken to be either the NG type or the Polyakov type. For the NG action, a gauge-independent formulation, similar to that adopted in \cite{hrt} for the string theory, was presented. The reparametrization invariances were manifested by the freedom in the choice of the multipliers enforcing the constraints of the theory. The implications of the boundary conditions in preserving the stability of the free membrane were discussed, highlighting the parallel with the string treatment. A set of quasi-orthonormal gauge fixing conditions was systematically obtained, which simplified the structure of the Hamiltonian.

A constrained analysis of the Polyakov action, contrary to the NG action,  led to the presence of second-class constraints. However, by an iterative prescription of computing Dirac brackets, the first-class sector was identified. The Dirac brackets of this sector were identical to the Poisson brackets and exactly matched with the involutive algebra found in the NG case. The analogue of the quasi-orthonormal gauge was also discussed in the Polyakov formulation. It naturally led to the choice of the metric which is used to perform calculations in the light-front variables \cite{wt}.
Moreover, in this gauge, the energy-momentum tensor was expressed as a combination of the constraints. On the constraint shell, this tensor was seen to have a vanishing trace.

A fundamental difference of the quasi-orthonormal gauge fixing in the two cases was pointed out. 
In the Polyakov case, gauge fixing entailed certain restrictions on the metric. Since the metric is regarded as an independent field, the gauge fixing does not affect the constraints of the theory which generate the reparametrization invariances. The discussion was thus confined to the Poisson algebra only. A similar gauge fixing in the NG case obviously restricts the target space coordinates. The first-class constraints get converted into second-class ones, thereby necessitating the use of Dirac brackets. Their evaluation is quite complicated due to non-linear terms. 

Since Dirac brackets were avoided in the Polyakov formulation, we proceeded to discuss noncommutativity only in this formulation. Also, cylindrical topology of the membrane was considered. We stress that, contrary to standard approaches \cite{aas1, aas2, ks, dmm, t, ch, rv},  boundary conditions were not treated as primary constraints of the theory. Our approach was in line with the previous treatment for string theory \cite{bcg}, which has been summarized in the appendix.
Thus, noncommutative algebra, if any, would be a manifestation of the Poisson brackets and not Dirac brackets. The noncommutative algebra was required to establish algebraic consistency of the boundary conditions with the basic Poisson brackets.
For the free theory it was
found that there was no clash between the boundary conditions and the Poisson brackets; hence there was no noncommutativity.

For the membrane interacting with a three-form potential a non-trivial algebraic relation was found that revealed the occurrence of noncommutativity, independent of any gauge choice or any approximations. Since this equation could not be solved, we passed on to its low-energy limit. Now this limit, which takes a membrane to a string, has been known for quite some time \cite{dhis} and has been studied or exploited in several circumstances \cite{l, bstt, s}. The cylindrical membrane is assumed to wrap around a circle, whose radius is taken to be vanishingly small. This enforces a double dimensional reduction with the eleven-dimensional compactified target space passing over to the ten-dimensional space while the membrane effectively reduces to an open string. We have studied this limit in some detail and showed  how the membrane boundary conditions, action
and the world-volume metric were transformed into the corresponding expressions for the string.
The equation governing noncommutativity in the membrane was likewise shown to reduce to the string example. 
 Since every point in D-brane can correspond to the end points of the 
cylindrical membrane, we get noncommutativity in D-brane coordinates also --- 
albeit in this low-energy limit. Of course, this feature of 
noncommutativity will persist even if this limit is not considered, otherwise the basic equation (\ref{514}) becomes inconsistent.

It might be worthwhile to pursue the connection of the low-energy limit with the noncommutativity. Indeed, instead of taking a vanishingly small radius, it is possible to retain terms up to some orders of the radius \cite{l}. This would presumably illuminate the nature of noncommutativity in the membrane (where it cannot be computed exactly) vis-\`a-vis the string.

%%%%%%%%%%%%%%%%%%%%%%%%%%%%%%%%%%%%%%%%%
\section*{Acknowledgements}
The authors thank Pradip Mukherjee for discussions. 
BC and KK gratefully acknowledge help from a stimulating discussion with
Koushik Ray.
RB thanks the Japan Society for Promotion of Science (JSPS) for support and
the members of the theory group, KEK, for their gracious hospitality.
KK  thanks the Council of Scientific and Industrial Research (CSIR),
Government of India, for financial support.

%%%%%%%%%%%%%%%%%%%%%%%%%%%%%%%%%%%%%%%%%%
\begin{appendix}

\section*{Appendix: Noncommutativity in Open String}
For an easy comparison of our results of open membrane with those
of open string, we summarize here the essential results of \cite{bcg}.

\subsection*{The Free Polyakov string}

The free Polyakov string action is
\begin{equation}
S_P = -{T_{S}\over 2} \int\!\!d\tau d\sigma{\sqrt{-g}}g^{ij}\partial_{i}X^{\mu}\partial_{j}X_{\mu}, \quad i,j = 0,1,
\label{901}
\end{equation}
where $T_{S}$ stands for string tension and $g_{ij}$, up to a Weyl factor,
is the induced metric $h_{ij}=\partial_{i}X^{\mu}\partial_{j}X_{\mu}$ on the
world-sheet. The canonical momenta are
\begin{equation}
\Pi_{\mu}= - T_{S}{\sqrt{-g}}\partial^{0}X_{\mu}, \quad 
\pi_{ij} = 0.
\label{903}
\end{equation}
It is clear that while $\Pi_{\mu}$ are genuine momenta, $\pi_{ij}\approx 0$
are the primary constraints of the theory. To determine the secondary
constraints, one can either follow the traditional Dirac's Hamiltonian
approach, or just read it off from the equation obtained by varying $g_{ij}$
since this is basically a Lagrange multiplier. This imposes the vanishing of
the symmetric energy-momentum tensor:
\begin{equation}
T_{ij}={2\over {\sqrt{-g}}}{\delta S_P\over \delta g^{ij}}
= - T_{S}\partial_{i}X^{\mu}\partial_{j}{X_{\mu}}
+ {T_{S}\over 2}g_{ij}g^{kl}\partial_{k}X^{\mu}\partial_{l}X_{\mu}
\approx 0.
\label{904}
\end{equation}
Because of the Weyl invariance, the energy-momentum tensor is traceless:
\begin{equation}
{T^i}_{i}=g^{ij}T_{ij} = 0
\label{905}
\end{equation}
so that only two components of $T_{ij}$ are independent. These components,
which are the constraints of the theory, are given by
\begin{eqnarray}
\chi_1\!\!&\equiv&\!\!2T_{S}gT^{00} = - 2T_{S}T_{11}
= \Pi^2+T_{S}^{2}h_{11}\approx 0,
\label{906} \\
\chi_2 \!\!&\equiv&\!\! {\sqrt{-g}}{T^0}_1 = \Pi_{\mu}\partial_{1}X^{\mu} \approx 0.
\label{907}
\end{eqnarray}
The canonical Hamiltonian density obtained from (\ref{901}) by a Legendre
transformation is given by
\begin{equation}
{\mathcal H}_{C} = {\sqrt{-g}}{T^0}_0
= {\sqrt{-g}\over 2T_{S}g_{11}}\chi_1+{g_{01}\over g_{11}}\chi_2,
\label{908}
\end{equation}
which, as expected, turns out to be a linear combination of the constraints.
The boundary condition written in terms of phase-space variables
is given by
\begin{equation}
\left[T_{S}\partial_1X^{\mu}+{\sqrt{-g}}g^{01}\Pi^{\mu}\right]_{\sigma=0,\pi} = 0,
\label{911}
\end{equation}
where the string parameters are in the region $-\infty \leq \tau \leq +\infty$,
$0\leq \sigma \leq \pi$. This boundary condition is incompatible with the first of the basic Poisson brackets:
\begin{eqnarray}
\left\{X^{\mu}(\tau,\sigma),\Pi_{\nu}(\tau,\sigma')\right\}
\!\!&=&\!\! \delta^{\mu}_{\nu}\delta(\sigma-\sigma'),
\label{912} \\
\left\{g_{ij}(\tau,\sigma),\pi^{kl}(\tau,\sigma')\right\}
\!\!&=&\!\! {1\over 2}(\delta^{k}_{i}\delta^{l}_{j}+ \delta^{l}_{i}\delta^{k}_{j})\delta(\sigma -\sigma').
\label{913}
\end{eqnarray}
From the basic Poisson brackets, it is easy to generate a first-class
(involutive) algebra:
\begin{eqnarray}
\left\{\chi _1(\sigma),\chi _1(\sigma')\right\}
\!\!&=&\!\! 4T_{S}^{2}\left[\chi _2(\sigma )+\chi _2(\sigma ')\right]\partial _{1}\delta(\sigma-\sigma'),
\nonumber \\
\left\{\chi _2(\sigma),\chi _1(\sigma')\right\}
\!\!&=&\!\! \left[\chi _1(\sigma)+\chi_1(\sigma')\right]\partial _{1}\delta(\sigma -\sigma'),
\label{914} \\
\left\{\chi_2(\sigma),\chi _2(\sigma')\right\}
\!\!&=&\!\! \left[\chi _2(\sigma)+\chi _2(\sigma')\right]\partial _{1}\delta(\sigma -\sigma').
\nonumber 
\end{eqnarray}
The constraints $\chi_1$ and $\chi_2$ generate the diffeomorphism
transformations.

The boundary condition~(\ref{911}) is not a constraint in
the Dirac sense, since it is applicable only at the boundary. Thus, there has
to be an appropriate modification in the Poisson brackets, to incorporate this
condition. This is not unexpected and occurs, for instance, in the example of
a free scalar field $\phi (x) $ in $1+1$ dimensions, subjected to periodic
boundary condition of period, say, $2\pi$.
There the Poisson bracket between the field $\phi(t,x)$ and its conjugate
momentum $\pi(t,x)$ is given by
\begin{equation}
\left\{\phi(t,x),\pi(t,y)\right\}= \delta_P(x-y),
\label{915}
\end{equation}
which is obtained
automatically if one starts with the canonical harmonic-oscillator algebra for
each mode in the Fourier space.

Before discussing the mixed condition~(\ref{911}), consider the simpler
Neumann-type condition.
Since the string coordinates $X^{\mu}(\tau ,\sigma)$ transform as a
world-sheet scalar under its reparametrization, it is more convenient to get
back to scalar field $\phi (t,x)$
defined on $(1+1)$-dimensional space-time, but with the periodic boundary
condition replaced by Neumann boundary condition,
\begin{equation}
\left.\partial_x\phi \right|_{\sigma=0,\pi} = 0,
\label{916}
\end{equation}
at the end points of a 1-dimensional box of compact size, i.e. of length $\pi$.
Correspondingly, the $\delta_P$ appearing there in the Poisson
bracket~(\ref{915}) --- consistent with periodic boundary condition --- 
has to be replaced now with a suitable ``delta function" incorporating
Neumann boundary condition, rather than periodic boundary condition.

Note that the following usual property of delta function is also
satisfied by $\delta_P(x-x')$:
\begin{equation}
\int_{-\pi}^{+\pi}dx'\delta_P(x-x')f(x') = f(x)
\label{917}
\end{equation}
for any periodic function $f(x)=f(x+2\pi)$ defined in the interval
$[-\pi, +\pi]$. Restricting to the case of even and odd functions,
$f_\pm(-x)=\pm f_\pm(x)$, the above integral reduces to
\begin{equation}
\int_0^{\pi}dx' \Delta_{\pm}(x,x')f_\pm(x') = f_\pm(x).
\label{918}
\end{equation}
Since any function $\phi(x)$ defined in the interval $[0,\pi ]$ can be
regarded as a part of an even or odd function $f_\pm(x)$ defined in the interval
$[-\pi,\pi]$, both $\Delta_{\pm}(x,x')$ act as delta functions
defined in half of the interval at the right i.e. $[0,\pi]$ as follows
from Eq.~(\ref{918}). It is still not clear which of these $\Delta_{\pm}(x,x')$
functions should replace $\delta_P(x-x')$ in the Poisson-bracket relation.
Now consider the Fourier decomposition of an arbitrary function
$f(x)$ satisfying periodic boundary condition $f(x)=f(x+2\pi )$:
\begin{equation}
f(x)=\sum_{n\in {\cal Z}}f_ne^{inx}.
\label{921}
\end{equation}
Clearly,
\[
f'(0) = i\sum_{n>0}n(f_n-f_{-n}), \quad
f'(\pi) = i\sum_{n>0}(-1)^nn(f_n-f_{-n}).
\]
Now for even and odd functions, the Fourier coefficients are related as
$f_{-n} =  \pm f_n$ so that Neumann boundary condition
$f'(0)=f'(\pi) = 0$ is satisfied if and only if $f(x)$ is even. Therefore,
one has to regard the scalar field $\phi(x)$ defined in the interval $[0,\pi ]$
and subjected to  Neumann boundary condition~(\ref{916}) as a part of an
even periodic function $f_+(x)$ defined in the extended interval
$[-\pi, +\pi]$. It thus follows that the appropriate Poisson bracket for the
scalar theory is given by
\[
\left\{\phi(t,x), \pi (t,x')\right\} = \Delta_{+}(x,x').
\]
It is straightforward to generalize it to the string case:
\begin{equation}
\left\{X^{\mu}(\tau,\sigma), \Pi_{\nu}(\tau,\sigma')\right\}
= \delta^{\mu}_{\nu}\Delta_{+}(\sigma,\sigma'),
\label{922}
\end{equation}
the Lorentz indices playing the role of ``isospin" indices, as viewed
from the world-sheet. Observe also that the other brackets
\begin{eqnarray}
\left\{X^{\mu}(\tau,\sigma),X^{\nu}(\tau,\sigma')\right\} \!\!&=&\!\! 0,
\label{923} \\
\left\{\Pi^{\mu}(\tau,\sigma), \Pi^{\nu}(\tau,\sigma')\right\} \!\!&=&\!\! 0
\label{924}
\end{eqnarray}
are consistent with the boundary conditions and hence remain unchanged.

The mixed condition~(\ref{911}) is
compatible with the modified brackets~(\ref{922}) and (\ref{924}), but not
with (\ref{923}). Therefore, make an ansatz:
\begin{equation}
\left\{X^{\mu}(\tau,\sigma),X^{\nu}(\tau,\sigma')\right\}
= C^{\mu\nu}(\sigma,\sigma'),
\label{925}
\end{equation}
where
\begin{equation}
C^{\mu\nu}(\sigma,\sigma') = -C^{\nu\mu}(\sigma',\sigma).
\label{926}
\end{equation}
Imposing the boundary condition~(\ref{911}) on this algebra, one gets
\begin{equation}
\left.\partial'_{1}C^{\mu\nu}(\sigma,\sigma')\right|_{\sigma'=0,\pi}
= \left.\partial_{1}C^{\mu\nu}(\sigma,\sigma')\right|_{\sigma=0,\pi}
= {\sqrt{-g}}g^{01}\eta^{\mu \nu}\Delta_{+}(\sigma,\sigma').
\label{927}
\end{equation}
For a restricted
class of metrics that satisfy $\partial_1g_{ij} = 0 $ it is possible to give a
quick solution of this equation as
\begin{equation}
C^{\mu\nu}(\sigma,\sigma')
= {\sqrt{-g}}g^{01}\eta^{\mu\nu}\left[\Theta(\sigma,\sigma')-\Theta(\sigma',\sigma)\right].
\label{928}
\end{equation}
This noncommutativity can
be made to vanish in gauges like conformal gauge, where $g^{01}=0$, thereby
restoring the usual commutative structure.
The essential structure of the involutive
algebra~(\ref{914}) is still preserved, only that $\delta(\sigma-\sigma')$
has to be replaced by $\Delta_+(\sigma,\sigma')$.

\subsection*{The Interacting Polyakov String}

The Polyakov action for a bosonic string moving in the presence of a constant
background Neveu-Schwarz two-form field $B_{\mu \nu}$ is given by
\begin{equation}
S_P = -\frac{T_{S}}{2}\int\!\! d\tau d\sigma \left[{\sqrt{-g}}g^{ij}{\partial}_iX^{\mu}{\partial}_jX_{\mu}
+ e\epsilon^{ij}B_{\mu\nu}\partial_iX^{\mu}\partial_jX^{\nu}\right],
\label{441}
\end{equation}
where a `coupling constant' $e$ has been introduced. A usual canonical
analysis leads to the following set of primary first-class constraints:
\begin{eqnarray}
gT^{00}={1\over 2}\left[(\Pi_{\mu}+eB_{\mu\nu}\partial_1X^{\nu})
(\Pi^{\mu}+eB^{\mu\nu}\partial_1X_{\nu})+T_{S}^{2}h_{11}\right]
\!\!&\approx&\!\! 0,
\label{442} \\
{\sqrt{-g}}{T^0}_1 = \Pi_{\mu}.\partial_1X^{\mu} \!\!&\approx&\!\! 0, 
\label{443}
\end{eqnarray}
where
\begin{equation}
\Pi_{\mu}=-T_{S}\left[{\sqrt{-g}}\partial^0X_{\mu} + eB_{\mu\nu}\partial_1X^{\nu}\right]
\label{444}
\end{equation}
is the momentum conjugate to $X^{\mu}$. The boundary condition
written in terms of phase-space variables is
\begin{equation}
\left[\partial_1X_{\mu} + \Pi^{\rho}(NM^{-1})_{\rho\mu}\right]_{\sigma=0,\pi}
= 0,
\label{448}
\end{equation}
where
\begin{equation}
{M^{\rho}}_{\mu}=T_{S}\left[{\delta^{\rho}}_{\mu}-e^2B^{\rho\nu}B_{\nu\mu}\right],
\quad
N_{\nu\mu}={g_{01}\over {\sqrt{-g}}}\eta_{\nu\mu}+eB_{\nu\mu}.
\label{447}
\end{equation}
The $\{X^\mu,\Pi_\nu\}$ Poissson bracket is the same as that of the free string
whereas considering the general structure~(\ref{925}) and exploiting the
above boundary condition, one obtains
\begin{equation}
\left.\partial_{1} C_{\mu\nu}(\sigma,\sigma')\right|_{\sigma=0,\pi}
= \left.(NM^{-1})_{\nu\mu}\Delta_+(\sigma,\sigma')\right|_{\sigma=0,\pi}.
\label{449}
\end{equation}
As in the free case, restricting to the class of metrics 
satisfying $\partial_{1}g_{ij}=0$, the above equation has a solution 
\begin{equation}
C_{\mu\nu}(\sigma,\sigma')
= {1\over 2}(NM^{-1})_{(\nu\mu )}\left[\Theta(\sigma,\sigma')-\Theta(\sigma',\sigma)\right]
+ {1\over 2}(NM^{-1})_{[\nu\mu]}\left[\Theta(\sigma,\sigma')+\Theta(\sigma',\sigma)-1\right].
\label{450}
\end{equation}
The modified algebra is gauge dependent; depends on the choice of the metric.
However, there is no choice for which the noncommutativity vanishes. To
show this, note that the origin of the noncommutativity is the presence of
non-vanishing $N_{\nu\mu }$ in the boundary condition~(\ref{448}). Vanishing
$N_{\nu\mu }$ would make $B_{\mu\nu}$ and $\eta_{\mu\nu}$ proportional which
obviously cannot happen, as the former is an antisymmetric and the latter is
a symmetric tensor. Hence, noncommutativity will persist for any choice of
world-sheet metric $g_{ij}$. Specially interesting are the
expressions for noncommutativity at the boundaries:
\begin{equation}
C_{\mu\nu}(0,0) = -C_{\mu\nu}(\pi,\pi) = - {1\over 2}(NM^{-1})_{[\nu\mu ]},
C_{\mu\nu}(0,\pi) = -C_{\mu\nu}(\pi,0) = -{1\over 2}(NM^{-1})_{(\nu\mu)}.
\label{451}
\end{equation}

\end{appendix}


\begin{thebibliography}{99}
\bibitem{jhwt} For a review of the theory of membranes, see
Jens Hoppe, {\tt hep-th/0206192} and \cite{wt}.
\bibitem{sw} N.~Seiberg and E.~Witten, JHEP {\bf 9909} (1999) 032
{\tt [hep-th/9908142]}.
%\bibitem{blt}E.~Bergshoeff, L.~London and P.~K.~Townsend, Class. Quant. Grav.
%{\bf 9} (1992) 2545 {\tt [hep-th/9206026]}.
\bibitem{bbss} E.~Bergshoeff, D.~S.~Berman, J.~P.~van der Schaar and P.~Sundell,
Nucl. Phys. {\bf B590} (2000) 173 {\tt [hep-th/0005026]}.
\bibitem{aas1} F.~Ardalan, H.~Arfaei and M.~M.~Sheikh-Jabbari, JHEP {\bf 9902} (1999) 016 {\tt [hep-th/9810072]}.
\bibitem{aas2} F.~Ardalan, H.~Arfaei and M.~M.~Sheikh-Jabbari, Nucl. Phys. {\bf B576} (2000) 578 {\tt [hep-th/9906161]}.
\bibitem{ks} S.~Kawamoto and N.~Sasakura, JHEP {\bf 0007} (2000)
014 {\tt [hep-th/0005123]}.
\bibitem{dmm} A.~Das, J.~Maharana and A.~Melikyan, JHEP {\bf 0104} (2001)
016 {\tt [hep-th/0103229]}.
\bibitem{t} Ken-Ichi Tezuka, Eur.~Phys.~J. {\bf C25} (2002) 465 {\tt [hep-th/0201171]}.
\bibitem{ch} Chong-Sun Chu and Pei-Ming Ho, Nucl. Phys. {\bf B550} (1999) 151 {\tt [hep-th/9812219]}.
\bibitem{rv} J.~M.~Romero and J.~D.~Vergara, {\tt hep-th/0212035}.
\bibitem{bcg} R.~Banerjee, B.~Chakraborty and S.~Ghosh, Phys. Lett. {\bf B537} 
(2002) 340 {\tt [hep-th/0203199]}.
\bibitem{hrt} A.~Hanson, T.~Regge and C.~Teitelboim, {\it Constrained 
Hamiltonian Systems}, Roma, Accademia Nazionale Dei Lincei (1976).
\bibitem{wt} W.~Taylor, Rev. Mod. Phys. {\bf 73} (2001) 419 {\tt [hep-th/0101126]}.
\bibitem{m} S.~Mandelstam, Phys.~Rept. {\bf 13} (1974) 259.
\bibitem{r} C.~Rebbi, Phys.~Rept. {\bf 12} (1974) 1.
\bibitem{gt} D.~M.~Gitman and I.~V.~Tyutin, {\it  Quantization of Fields with Constraints,} Springer-Verlag, 1990.
\bibitem{l} U.~Lindstrom, Phys.~Lett.~{\bf B218} (1989) 315.
\bibitem{dhis} M.~J.~Duff, Paul S.~Howe, T.~Inami and K.~S.~Stelle, Phys.~Lett.~{\bf B191} (1987) 70.
\bibitem{bstt} E.~Bergshoeff, E.~Sezgin and P.~K.~Townsend, Annals Phys. {\bf 185} (1988) 330.
\bibitem{s} L.~Smolin, Phys.~Rev.~{\bf D57} (1998) 6216 {\tt [hep-th/9710191]}.   
\end{thebibliography}
\end{document}